\documentclass[aps, prc, 10pt, tightenlines, letterpaper, amsmath, amssymb, preprintnumbers, floatfix, longbibliography, nofootinbib,superscriptaddress]{revtex4-2}
\usepackage[T1]{fontenc}
\usepackage{bm}
\usepackage{xspace}
\usepackage[dvipsnames]{xcolor}
\usepackage{graphicx}
\usepackage{tabularx}
\usepackage{enumitem}
\usepackage{qcircuit}
\usepackage{braket}
\usepackage{dsfont}
\usepackage{placeins}
\usepackage[normalem]{ulem}
\usepackage{diagbox}
\usepackage{soul}
\newcolumntype{C}[1]{>{\centering\let\newline\\\arraybackslash\hspace{0pt}}m{#1}}

\usepackage{amsmath, nccmath}
\usepackage{lipsum}
\usepackage{cancel}
\usepackage{bm}

\usepackage[hypertexnames=false]{hyperref}
\hypersetup{
    colorlinks=true,       
    linkcolor=blue,          
    citecolor=blue,        
    filecolor=blue,      
    urlcolor=blue           
}

\newcolumntype{Y}{>{\centering\arraybackslash}X}

\usepackage{mathtools}
\usepackage{orcidlink}

\definecolor{lavenderindigo}{rgb}{0.58, 0.34, 0.92}

\usepackage{stackengine}
\stackMath
\newcommand\tenq[2][1]{%
 \def\useanchorwidth{T}%
  \ifnum#1>1%
    \stackunder[0pt]{\tenq[\numexpr#1-1\relax]{#2}}{\scriptscriptstyle\sim}%
  \else%
    \stackunder[1pt]{#2}{\scriptscriptstyle\sim}%
  \fi%
}

\DeclareMathOperator{\sech}{sech}

\begin{document}

\begin{figure}
\vskip -1.cm
\leftline{\includegraphics[width=0.15\textwidth]{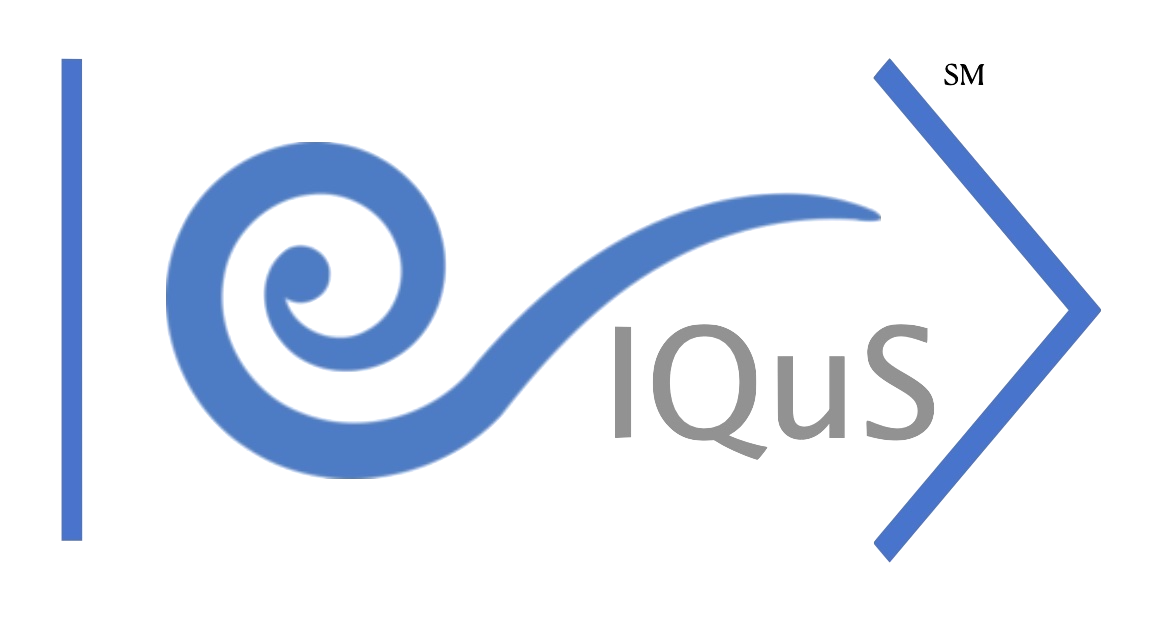}}
\vskip -0.5cm
\end{figure}

\title{Steps Toward Quantum Simulations of Hadronization and Energy-Loss \texorpdfstring{\\}{} in Dense Matter}

\author{Roland C.~Farrell\,\orcidlink{0000-0001-7189-0424
}}
\email{rolanf2@uw.edu}
\affiliation{InQubator for Quantum Simulation (IQuS), Department of Physics, University of Washington, Seattle, WA 98195, USA.}
\affiliation{Albert Einstein Center for Fundamental Physics, Institut für Theoretische Physik,
Universität Bern, Sidlerstrasse 5, CH-3012 Bern, Switzerland}
\author{Marc Illa\,\orcidlink{0000-0003-3570-2849}}
\email{marcilla@uw.edu}
\affiliation{InQubator for Quantum Simulation (IQuS), Department of Physics, University of Washington, Seattle, WA 98195, USA.}
\author{Martin J.~Savage\,\orcidlink{0000-0001-6502-7106}}
\email{mjs5@uw.edu}
\thanks{On leave from the Institute for Nuclear Theory.}
\affiliation{InQubator for Quantum Simulation (IQuS), Department of Physics, University of Washington, Seattle, WA 98195, USA.}

\preprint{IQuS@UW-21-076, NT@UW-24-06}
\date{\today}

\begin{abstract}
\noindent
A framework for simulating the real-time dynamics of
composite
particles in a simple model of dense matter that is amenable to quantum computers is developed.
As a demonstration, we perform
classical simulations of heavy-hadrons propagating through a dense medium in the Schwinger model.
Measurements of the time-dependent energy and charge density are used to identify mechanisms responsible for energy loss and hadron production (hadronization).
A study of entanglement dynamics highlights the importance of quantum coherence between the particles that make up the dense medium.
Throughout this work,
care is taken to isolate, and remove, phenomena that 
arise solely from a finite lattice spacing.
It is found that signatures of entanglement are 
more sensitive to lattice 
artifacts
than other observables.
Toward quantum simulations, we present  an efficient method and the corresponding quantum circuits for preparing ground states in the presence of heavy mesons.
These circuits are used to estimate the resources required to simulate in-medium energy loss and hadronization in the Schwinger model 
using quantum computers.
\end{abstract}

\maketitle
\newpage{}
\tableofcontents
\newpage{}

\section{Introduction}
\label{sec:intro}
\noindent
An improved understanding of the transport of energy, momentum, flavor, and other quantum numbers 
in non-equilibrium strongly-interacting dense matter is needed to  
refine predictive capabilities for 
the extreme matter created in the early universe and in astrophysical environments.
Motivation and inspiration comes from ongoing and planned experiments of heavy-ions collisions~\cite{Dusling:2015gta,Busza:2018rrf} and from astronomical observations of multi-messenger signals~\cite{LIGOScientific:2017ync,Troja:2018uns,KAGRA:2021vkt,LIGOScientific:2020aai}.
The state-of-the-art predictions for the structure and dynamics of extreme matter 
integrate the experimental results from these programs with analytic and computational techniques, 
and phenomenological modeling~\cite{Lovato:2022vgq,Fevre:2023swv,Jacobi:2023olu,Brandes:2023hma}.

Knowledge of energy-loss and stopping ranges of electrically charged particles, from electrons to nuclei, in ordinary materials is essential to many scientific, technological, societal, and therapeutic endeavors, including in protecting humans and scientific equipment from cosmic rays and background radiation.
It is also critical to designing ultra-sensitive experiments to search for new physics that interacts weakly with ordinary matter, such as searches for Dark Matter~\cite{Trickle:2019nya,Baxter:2022dkm} and $0\nu\beta\beta$-decay of nuclei~\cite{Giuliani:2012zu,Dolinski:2019nrj}.
Long-term experimental programs have measured the energy loss for particles penetrating a wide selection of materials over a broad spectrum of energies.
Theoretical descriptions of the underlying mechanisms are well-established for electrically-charged particles and $\gamma$-rays over a significant energy regime, including, for example, the effects of elastic and inelastic collisions, pair-production and bremsstrahlung radiation~\cite{Tsai:1973py,SELTZER1984665}.
Interactions with nuclei at higher energies remain the focus of work at current and future high-energy colliders and fixed-target experiments~\cite{Meehan:2017cum,Maurice:2017iom,Hadjidakis:2018ifr,Barschel:2020drr,AbdulKhalek:2021gbh}, e.g., experiments at the Thomas Jefferson National Accelerator Facility (TJNAF), Brookhaven National Laboratory (BNL), or at the Large Hadron Collider (LHC).
Together, experiment, theory, and computation have established an extensive catalogue enabling predictions for the behavior of charged particles moving through a variety of materials (a brief summary can be found in the Particle Data Group~\cite{Workman:2022ynf}).
In contrast, energy loss mechanisms in dense nuclear matter in (and out of) equilibrium are less well understood. 
There has been significant theoretical progress on this topic 
(e.g., see Ref.~\cite{zhao2023hadronization}), 
largely driven by input from heavy-ion collision experiments at BNL~\cite{PHENIX:2004vcz,BRAHMS:2004adc,PHOBOS:2004zne,STAR:2005gfr,STAR:2010vob,STAR:2017sal} and the LHC~\cite{Loizides:2016tew,Foka:2016zdb,Foka:2016vta}.
As an example, because of their mass and compactness, the transport properties, yields, and distributions of 
heavy quarks and
quarkonia systems provide insights into the nature of the  matter created in heavy-ion collisions (for recent reviews on different theoretical approaches, see Refs.~\cite{Rothkopf:2019ipj,Akamatsu:2020ypb,Chapon:2020heu,Yao:2021lus,Montana:2023sft}).
However, in processes where quantum coherence and entanglement play a significant role, conventional methods, e.g. Monte-Carlo, experience limitations. 
In addition to the ongoing formal and numerical efforts to quantify these effects in heavy-ion collisions, e.g., Refs.~\cite{Casalderrey-Solana:2012evi,Arnold:2024bph}, 
quantum simulations are being developed to study the role of quantum coherence in parton showers~\cite{Bauer:2019qxa,Bepari:2020xqi,Bepari:2021kwv,Macaluso:2021ngq,Chigusa:2022act,Gustafson:2022dsq,Bauer:2023ujy} and jets in medium~\cite{Barata:2021yri,Yao:2022eqm,Barata:2022wim,Barata:2023clv,Du:2023ewh}.

Simulations of non-perturbative phenomena in QCD are typically performed using the framework of lattice gauge theory.
Lattice QCD simulations using classical computers have been enormously successful in determining static properties (e.g., see Ref.~\cite{FlavourLatticeAveragingGroupFLAG:2021npn} for a review of flavor physics in the meson sector, or Ref.~\cite{Davoudi:2020ngi} for a review in the multi-nucleon sector), but typically break down for simulations of real-time dynamics or with finite densities because of sign problems in the Monte-Carlo sampling.
In contrast, it is believed that such simulations can be performed efficiently
using quantum computers.
Limited by the capabilities of available quantum hardware, software, and algorithms, 
current research has focused on small-scale simulations of model lattice gauge theories.
This includes a number of early small-scale 
simulations using noisy intermediate-scale quantum (NISQ)~\cite{Preskill:2018jim} 
era quantum computers of low dimensional lattice gauge theories~\cite{Martinez:2016yna,Klco:2018kyo,Kokail:2018eiw,Lu:2018pjk,Klco:2019evd,Mil:2019pbt,Yang:2020yer,Ciavarella:2021nmj,Bauer:2021gup,Atas:2021ext,ARahman:2021ktn,Mazzola:2021hma,deJong:2021wsd,Gong:2021bcp,Zhou:2021kdl,Riechert:2021ink,Ciavarella:2021lel,Nguyen:2021hyk,PhysRevResearch.5.023010,Illa:2022jqb,Mildenberger:2022jqr,ARahman:2022tkr,Asaduzzaman:2022bpi,Farrell:2022wyt,Atas:2022dqm,Farrell:2022vyh,Mueller:2022xbg,Pomarico:2023png,Charles:2023zbl, zhang2023observation, Ciavarella:2023mfc,Meth:2023wzd,Borzenkova:2023xaf,Schuster:2023klj,Angelides:2023noe,Kavaki:2024ijd,Davoudi:2024wyv,Turro:2024pxu,Ciavarella:2024fzw}, 
and recent larger-scale simulations of 1+1D quantum electrodynamics (QED)~\cite{Farrell:2023fgd,Farrell:2024fit}, the Schwinger model~\cite{Schwinger:1962tp}.
Like QCD, the Schwinger model is a confining gauge theory and is a popular test-bed for new ideas.
Relevant to the present work are previous classical simulations of jet and hadron production~\cite{Li:2020uhl,Bauer:2021gup,Barata:2021yri,Li:2021zaw,Barata:2022wim,Florio:2023dke,Barata:2023clv,Florio:2024aix,Yao:2022eqm}, string breaking~\cite{Hebenstreit:2013baa,Pichler:2015yqa,Kasper:2015cca,Sala:2018dui,Magnifico:2019kyj,Lee:2023urk,Belyansky:2023rgh,Barata:2023jgd}, and of the associated entanglement evolution~\cite{Chai:2023qpq,Papaefstathiou:2024zsu}.

In this work, classical simulations of the Schwinger model are performed to determine the energy-loss and other observables associated with particles moving through dense matter.
An extensive study of heavy-hadrons moving through the 
lattice vacuum is performed to
isolate lattice artifacts that arise from the breaking of Lorentz symmetry, 
and will not survive in the continuum.
These lattice artifacts are magnified in certain entanglement measures, 
and lead to energy loss and light-hadron production 
even for a heavy-hadron moving through the vacuum at constant velocity.
Once parameters are found where these artifacts are minimized, the propagation of heavy-hadrons through a medium of static heavy-hadrons is considered.
Energy loss due to the production of hadrons (hadronization) and internal excitations are identified, and the crucial role of quantum coherence is emphasized.
These classical simulations are limited to system sizes of $L=12$ spatial sites (24 staggered sites), and we present scalable quantum circuits for state preparation and estimate the resources required for large-scale quantum simulations of these phenomena.

\section{The Simulation Strategy}
\label{sec:framework}
\noindent
To investigate how charges move through dense matter,
heavy ``external'' charges are introduced into the Schwinger model, 
with positions specified by classical trajectories.
Heavy charges with fixed positions are used to model a medium, with 
a density controlled by the separation between the heavy charges,
which in these calculations are in units of the lattice spacing. Additional heavy charges moving across these regions probe energy-loss, 
fragmentation, hadronization, and entanglement arising from propagation through a dense medium. 
These heavy charges emulate the heavy fields 
in heavy-quark effective theory (HQET)~\cite{Isgur:1989vq,Isgur:1990yhj,Eichten:1989zv,Georgi:1990um,Grinstein:1990mj} that are used to define 
a systematic expansion about the heavy-quark limit.
In this limit, analogous to a B-meson,
a heavy-hadron in the Schwinger model is composed of a single heavy charge 
that is electrically neutralized by a ``cloud'' of light charges.
Important to the current treatment
is that the position, velocity, and acceleration of the moving heavy charge 
are well-defined throughout its motion.
To access the desired physics, we
choose a classical trajectory where the heavy charge accelerates to a constant velocity, 
moves through the dense medium, and then decelerates to rest.

\subsection{The Lattice Schwinger Model Hamiltonian}
\label{sec:SM}
\noindent
The fields in the Schwinger model
are constrained by Gauss's law such that the photon field can be completely specified by the 
configuration of electric charges, and has no independent dynamics.
The photon can be removed as an explicit degree of freedom, 
and the 
contributions from the fermion-photon interactions are included via
a linear Coulomb potential between the fermions.
The Kogut-Susskind (KS) staggered discretization of the fermion fields is used 
to define the Hamiltonian~\cite{Kogut:1974ag,Banks:1975gq}
in our development of the relevant quantum simulations.
In axial gauge,
with open boundary conditions (OBCs),
zero background electric field, 
and mapping fermions to qubits with the Jordan-Wigner transformation,
the lattice Hamiltonian for a single fermion flavor, 
given in Refs.~\cite{Banks:1975gq,Kokail:2018eiw}, is
\begin{align}
\hat H & 
\ =\  \hat H_m + \hat H_{kin} + \hat H_{el} 
\ = \ \frac{m}{ 2}\sum_{j=0}^{2L-1}\ \left[ (-1)^j \hat Z_j + \hat{I} \right] 
\ + \ \frac{1}{2}\sum_{j=0}^{2L-2}\ \left( \hat \sigma^+_j \hat\sigma^-_{j+1} + {\rm h.c.} \right) 
\ + \ \frac{g^2}{ 2}\sum_{j=0}^{2L-2}\bigg (\sum_{k\leq j} \hat{q}_k +Q_k \bigg )^2 
\ ,
\nonumber \\ 
\hat q_k & \ = \ -\frac{1}{2}\left[ \hat Z_k + (-1)^k\hat{I} \right] 
\ ,
\label{eq:Hgf}
\end{align}
where $L$ is the number of spatial sites, corresponding to $2L$ staggered sites, with electrons (positrons) on even (odd) numbered sites.
The electric charge operator $\hat q_k$ acts on the $k^{\rm th}$ staggered site, and $Q_k$ is a heavy background charge.
The heavy charges have been included as discontinuities in Gauss's law, which
(on the lattice) is,
\begin{equation}
{\bf E}_k - {\bf E}_{k-1} = q_k + Q_k \ ,
\end{equation}
where ${\bf E}_k$ is the electric field on the link between staggered sites $k$ and $k+1$.
The  input lattice mass of the electron is $m$, electric charge is $g$, and $\hat \sigma^\pm = \frac{1}{2}(\hat X \pm i \hat Y)$ and $\hat Z$ are Pauli operators.
Due to confinement and without background charges, the low-energy excitations are charge-neutral bound states of electrons and positrons. 
The hadron mass $m_{\text{hadron}}$ and confinement length
$\xi \sim m_{\text{hadron}}^{-1}$ depend non-perturbatively on $m$ and $g$.
The range of $\{m,g\}$ values used in this work give rise to $1.3 \leq m_{\text{hadron}}^{-1} \leq 1.8$ 
that is well contained inside of the volumes accessible to classical simulation, 
$\xi \ll L$.
All lengths are measured in units of the staggered lattice spacing that has been set to
$a_{\text{staggered}} = a_{\text{spatial}}/2=1$.
The conserved quantities and symmetries of this system are total electric charge 
$\hat{q}_{tot} = \sum_{k} \hat{q}_k$, 
time reversal, and
the combined operation of charge conjugation and parity (CP).\footnote{The CP symmetry is realized in the $q_{tot}=0$ sector with no external charges as the composition of a spin-flip and a reflection through the mid-point of the lattice.}
It is important to keep in mind that
this Hamiltonian measures the energy in the light degrees of freedom, as there is no mass 
or kinetic term for the heavy charges.
An explicit expression for $\hat{H}_{el}$ with the charge operator $\hat{q}_k$ expanded in terms of $\hat{Z}_k$ can be found in App.~\ref{app:hamZ}.

The electric interaction has ${\cal O}(L^2)$ terms, with coefficients that grow with increasing separation of the charges due to the linear Coulomb potential.
For use in quantum simulation, the evolution of the electric interaction $e^{- i t\hat{H}_{el}}$ is converted into a sequence of gates. 
This requires $R_{ZZ}$ gates between all pairs of qubits, with a total of ${\cal O}(L^2)$ gates and a circuit depth that is ${\cal O}(L)$ with all-to-all connectivity between qubits, 
or ${\cal O}(L^2)$ with nearest-neighbor connectivity.
This scaling renders large-scale quantum simulations with this Hamiltonian impractical in the 
near-term,
and violates one of the criteria required for efficient simulations at scale~\cite{DiVincenzo:2000tra}.
Fortunately, the Schwinger model is confining at low temperatures and without a $\theta$-term, 
and charges are screened beyond a confinement length.
In recent work~\cite{Farrell:2024fit}, 
we showed that confinement can be used to re-organize the interactions 
in a way that allows for a truncation of terms $\hat{q}_k \hat{q}_{k+d}$ 
that are separated by $d\gtrsim\xi$, with
errors that are exponentially suppressed by $\sim e^{-d/\xi}$.
Time evolution using this truncated interaction only requires $\mathcal{O}(\xi L)$ gates with a circuit depth of ${\cal O}(\xi)$ 
with all-to-all connectivity, 
and ${\cal O}(\xi^2)$ with nearest-neighbor connectivity. 
This truncation, while different in nature, is similar to the cutoff in the maximum value of the electric flux used in tensor network calculations of the Schwinger model~\cite{Banuls:2015sta,Banuls:2016lkq,Papaefstathiou:2024zsu}.

In the vacuum, two-point correlations between charges on distant staggered lattice sites
can be large due to disconnected contributions,
\begin{align}
\langle \hat{q}_k \hat{q}_{k+d}\rangle \ \sim \ e^{-d/\xi} \ + \ \langle \hat{q}_k \rangle\langle \hat{q}_{k+d}\rangle \ .
\end{align}
The $\langle \hat{q}_k \rangle$ are negative (positive) on electron (positron) sites and prevent convergence of the truncated interaction.
This problem is solved by
performing a change of variables to total charges and dipole moments defined on spatial sites~\cite{Farrell:2024fit},
\begin{equation}
\hat{\overline{q}}_n \ = \ \hat{q}_{2n} + \hat{q}_{2n+1} + Q_{2n} + Q_{2n+1} \ , \quad \hat{\delta}_n \ = \ \hat{q}_{2n} - \hat{q}_{2n+1} + Q_{2n} - Q_{2n+1} \ .
\end{equation}
Note that both the heavy and light charges have been absorbed into these definitions.
Charges on spatial sites satisfy $\langle \hat{\overline{q}}_n \rangle = 0$ 
away from the boundaries and heavy-charges. 
Near the heavy-charges, this is modified due the finite size of the screening ``cloud''. 
However, 
averaging over the extent of the screening 
will restore this property, and correlations between well-separated spatial charges 
(averaged over the screening) will decay exponentially.

\subsection{Heavy-\texorpdfstring{$Q$}{} Trajectories}
\label{sec:trajsQ}
\begin{figure}[!t]
\centering
\includegraphics[width=\textwidth]{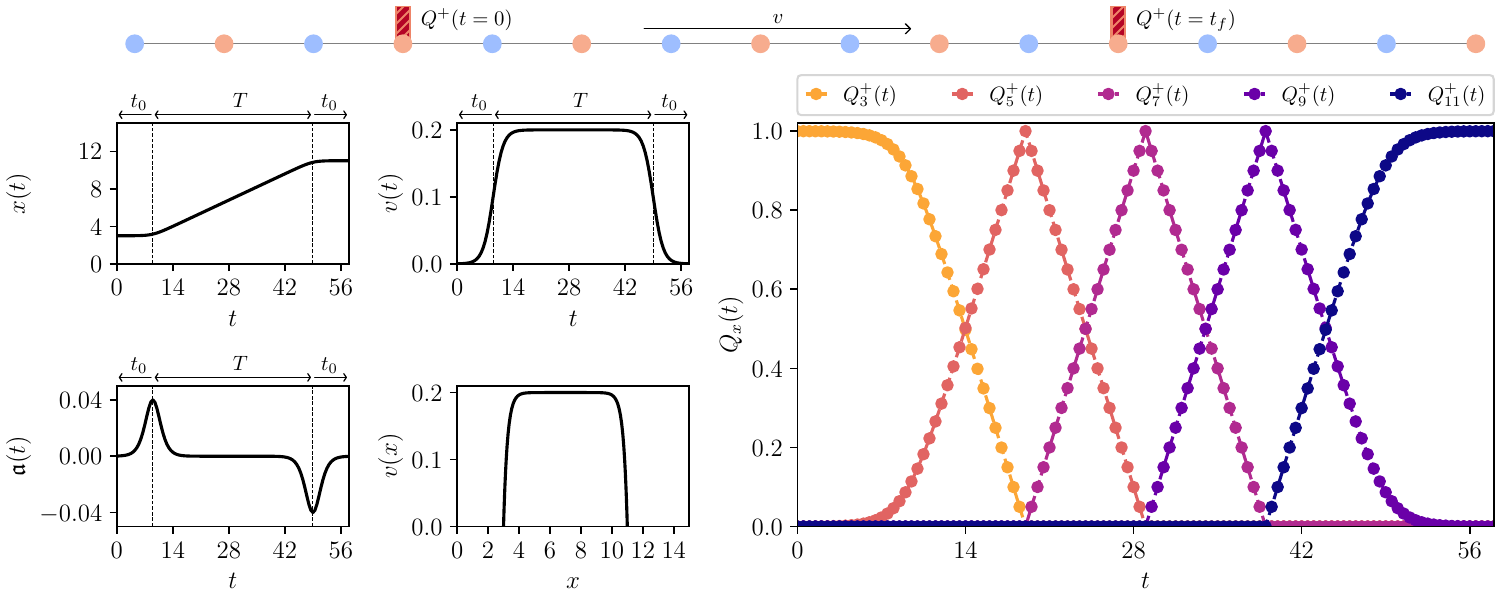}
\caption{
An example of a
trajectory 
moving a heavy-$Q^+$ from $x_0=3$ to $x_f=11$ on a $L=8$ lattice.
The panels on the left show
$x(t)$, $v(t)$, $\mathfrak{a}(t)$, and $v(x)$ for
$v_{\rm max}=0.2$ and $\mathfrak{a}_{\rm max}=v_{\rm max}/5$,
using
$t_0=9$ and $T=40$, as defined in Eq.~\eqref{eq:xva_softer_rels}. 
The panel on the right shows 
the distribution of the heavy charge between the two nearest positron sites
as a function of time, 
computed via the procedure described in Eq.~\eqref{eq:chargePart}.
}

\label{fig:TRAJv0p2a0p2}
\end{figure}

\noindent
In our treatment, moving heavy-$Q^+$s follow
 a trajectory with
a smooth acceleration up to a uniform velocity, 
followed by a smooth deceleration to rest.
A trajectory is chosen that minimizes the maximum acceleration, thus reducing the energy loss associated with an accelerating charge.
The physics should only depend on the uniform velocity of the heavy charge, leaving a lot of freedom in how the equations of motion of the trajectory are parameterized. One choice is,
\begin{align}
x(t) & \ =
\ \frac{v_{\text{max}}^2}{4 {\mathfrak a}_{\text{max}}}\ 
\log \frac{\cosh \left[ \beta (t-t_0) \right]}{\cosh \left[ \beta (t-t_0-T) \right]} 
\ + \ \frac{x_f + x_0}{2}
\ ,
\nonumber\\[4pt]
v(t) & \ = \  \frac{v_{\text{max}}}{2}\ \left( 
\tanh \left[ \beta (t-t_0) \right] -  \tanh \left[ \beta (t-t_0-T)  \right] 
\right)
\ ,
\nonumber\\[4pt]
{\mathfrak a}(t) & \ = \ {\mathfrak a}_{\text{max}} \  \left( 
\sech^2 \left[ \beta (t-t_0) \right] -  \sech^2 \left[ \beta (t-t_0-T)  \right] 
\right)
\ ,
\label{eq:xva_softer}
\end{align}
where 
\begin{equation}
\beta\ =\ 2\frac{{\mathfrak a}_{\rm max}}{v_{\text{max}}}
\ ,\quad 
t_0\ =\ \left\lfloor \frac{\text{arccosh}[\sqrt{{\mathfrak a}_{\text{max}}/{\mathfrak a}(0)}]}{\beta} \right\rceil
\ ,\quad 
T\ =\ \frac{x_f-x_0}{v_{\text{max}}}
\ .
\label{eq:xva_softer_rels}
\end{equation}
The trajectories are defined by the maximum velocity ($v_{\text{max}}$) and acceleration ($\mathfrak{a}_{\text{max}}$), as well as the initial position
($x_0$) and final position ($x_f$) of the heavy charge. 
The variable $t_0$ is fixed by the initial acceleration ${\mathfrak a}(0)$, which for our numerical studies is set to ${\mathfrak a}(0) = 10^{-4}$, and 
$\lfloor \cdot\cdot\cdot \rceil$ denotes the round function to the nearest integer.
The continuous position $x(t)$ is distributed among the two nearest odd numbered (positron) staggered sites to match the positive heavy charge.
Defining $x_{q1}$ to be the smaller numbered site and $x_{q2}=x_{q1+2}$ to be the larger numbered 
site, the charge is distributed as,
\begin{equation}
Q_{q1} \ = \ 
\frac{Q}{2} \left(x_{q2}-x(t)\right)
\ , \quad 
Q_{q2} \ = \ 
\frac{Q}{2}  \left(x(t)-x_{q1}\right)
    \ ,
\label{eq:chargePart}
\end{equation}
where $Q$ is the value of the heavy charge.
To minimize boundary effects, the initial and final positions of a heavy-$Q^+$ are placed as far as practical from the edges of the lattice, but in such a way to have an extended period of constant velocity. 
An example trajectory is shown in Fig.~\ref{fig:TRAJv0p2a0p2},
where a heavy-$Q^+$ moves
from $x_0=3$ to $x_f=11$ with the constraints that $v_{\rm max}=0.2$ and ${\mathfrak a}_{\rm max}=0.04$.
Throughout this work, we will set ${\mathfrak a}_{\rm max}=v_{\rm max}/5$, 
which we have verified does not lead to an appreciable energy loss due to the 
radiation from an accelerating charge.
%

\subsection{Maximum Lattice Velocity}
\label{sec:DispRels}
\noindent
Features of the dynamical simulations that will be
presented in the following sections 
can be understood from the lattice dispersion relations.
The spectrum of the Schwinger model has been extensively studied in the literature, 
with a focus on the first excitations in the charge $q_{tot}=0$ sector (scalar and vector mesons)~\cite{Hamer:1982mx,Berruto:1997jv,Sriganesh:1999ws,Byrnes:2002nv,Cichy:2012rw,Banuls:2013jaa}.
It is convenient to first consider the theory with $g=0$, 
corresponding to non-interacting electrons and positrons.
The lattice dispersion relation for the electrons (in the charge $q_{tot}=-1$ sector)
resulting from the Hamiltonian given in Eq.~\eqref{eq:Hgf}, 
subject to OBCs, is
\begin{equation}
    E^2 \ = \ m^2 + \sin^2\left( K/2 \right) 
    \ ,\ \ 
    K \ = \ \frac{(\overline{n}+\frac{1}{2}) \pi}{L+\frac{1}{2}}
    \ ,
\label{eq:dispF}
\end{equation}
where $\overline{n}=\{0,1,\ldots, L-1\}$.
These energies are the gap above the vacuum, and
there are more energy levels in the $g=0$ spectrum beyond single-particle states
corresponding to multi-particle excitations.
The form of this (electron) dispersion relation is from, in part, the spatial lattice spacing, relevant for the distance between adjacent lattice sites, being twice the staggered lattice spacing, $a_{\text{staggered}}=1$ and $a_{\text{spatial}}=2$. 
The electron dispersion is relevant for simulations with a moving heavy-$Q^+$, whose charge is neutralized by 
electrons.
The group velocity of electrons is  
\begin{equation}
v \ = \ \frac{dE}{dK} \ = \ \frac{\sin K}{4\sqrt{m^2 + \sin^2 \left (K/2 \right )}} \ ,
\label{eq:dispBv}
\end{equation}
where the limit $L\to\infty$ has been taken in order
to define the derivative.
Unlike in the continuum, there is a maximum group velocity, 
\begin{equation}
v_\star \ = \ \frac{1}{2}\left (\sqrt{m^2+1}-m \right )
    \ ,
\label{eq:Vstar}
\end{equation}
which reduces to the speed of light, $c=\frac{1}{2}$ 
(in spatial lattice units),
in the $m\rightarrow 0$ (continuum) limit. 
This maximum velocity will persist in the interacting theory, with a value that is shifted away from $v_{\star}$.
Therefore, on the lattice, there is a critical velocity of the heavy charge that exceeds the maximum group velocity of the light degrees of freedom. 
This critical velocity is lower than the speed of light, which is 1 staggered site per unit time.
This indicates that particle production can occur 
on the lattice even when the heavy charge is moving at constant velocity
because some or all of the light degrees of freedom are unable to ``keep up'' with the charge for sufficiently high velocity.
Conceptually, the moving heavy charge will separate from the light degrees of freedom that were initially screening it, exposing the vacuum to an electric field, 
which will create hadrons in its wake.
When determining the energy loss of a 
charged particle
in medium at a non-zero lattice spacing, 
the energy loss into the vacuum will be subtracted.

\section{Classical Simulations}
\label{sec:CSims}
\noindent
Classical simulations of a selection of heavy-$Q^+$ trajectories with different ``dense mediums'' are performed.
These simulations determine the state at time $t$, $\vert \psi(t) \rangle$, from 
the ground  state in a particular charge sector,
$\vert \psi(0) \rangle$,
via a Trotterized time evolution associated with the time-dependent Hamiltonian,
\begin{align}
\vert \psi(t) \rangle \ = \ \mathcal{T} \prod_{j=1}^{t/\Delta t}\, e^{- i \, \Delta t \, \hat{H}(j \, \Delta t)} \vert \psi(0) \rangle \ ,
\label{eq:tevol}
\end{align}
where $\mathcal{T}$ denotes the time-ordered product.
It was found that a (minimal) time step of $\Delta t=0.25$ was sufficient for the convergence of the observables considered.\footnote{For small values of $v$, the time step $\Delta t$ can be increased. Explicitly, for $v\leq 0.05$, $\Delta t=2.0$; for $0.05 < v\leq 0.1$, $\Delta t=1.5$; for $0.1 < v\leq 0.2$, $\Delta t=1.0$; for $0.2 < v\leq 0.4$, $\Delta t=0.5$; for $0.4 < v\leq 0.99$, $\Delta t=0.25$.} 
The time-dependent energy, charge distribution and various entanglement measures 
are determined from $\vert \psi(t) \rangle$.

\subsection{A Heavy-\texorpdfstring{$Q^+$}{} Moving Through the Lattice Vacuum}
\label{sec:HQVAC}
\noindent
Lorentz invariance is broken down to discrete translational invariance
in simulations using a spatial lattice.
The lattice dispersion relation
allows processes to occur that  are forbidden in the continuum by energy-momentum conservation, 
such as pair production below threshold.
This leads to {\it increasing energy} in the light degrees of freedom
as a heavy-$Q^+$ moves with constant velocity across the lattice vacuum,
which we connect to the more standard framework of {\it energy-loss} by the moving heavy-$Q^+$.
The workflow that we employ to simulate the dynamics of
a neutralized heavy-$Q^+$ moving across the lattice vacuum is the following:
\begin{enumerate}
    \item 
    Determine the vacuum state, $|\psi_{\rm vac}\rangle$,  
    and low-lying excited states without background charges.
    This defines the vacuum energy ($E_{\rm vac}$), the mass of the hadronic excitations
    in the light sector, the chiral condensate, and other vacuum observables.
    \item 
    Determine the ground state with a neutralized
    heavy-$Q^+$ at rest at site $x_0$, $|\psi_{\rm vac}\rangle_{Q^+_{\{x_0\}}}$. 
    The energy gap above $E_{\rm vac}$ defines the mass of the heavy hadron 
    (analogous to the B-meson), or more precisely the lattice evaluation of $\overline{\Lambda}$ in HQET~\cite{Falk:1992ws,Luke:1993za,Kronfeld:2000gk,Gambino:2017vkx,FermilabLattice:2018est}.
     \item 
    Time evolve the state $|\psi (0)\rangle=|\psi_{\rm vac}\rangle_{Q^+_{\{x_0,v\}}}$ using Eq.~\eqref{eq:tevol}, with the heavy-$Q^+$ trajectory, $x(t)$, defined above in Eqs.~\eqref{eq:xva_softer} and \eqref{eq:chargePart}.
    At each time step, the relevant observables are computed, 
    including the total energy, given in Eq.~\eqref{eq:Etottx}.
\end{enumerate}
It is convenient to define observables as functions of the position of the moving heavy-charge, instead of time. For example, the total energy is
\begin{equation}
E_{Q^+}(x) \ = \ \langle \  \psi[t(x)] \  \vert  \ \hat{H}[t(x)] \  \vert \  \psi[t(x)] \ \rangle_{Q^+_{\{x_0,v\}}} 
\ ,
\label{eq:Etottx}
\end{equation}
where $t(x)$ can be determined from inverting the heavy-$Q^+$ trajectory $x(t)$ from Eq.~\eqref{eq:xva_softer}. 
All the quantities displayed
in the rest of the paper will depend on the position of the heavy-$Q^+$.

\begin{figure}[!ht]
\centering
\includegraphics[width=\textwidth]{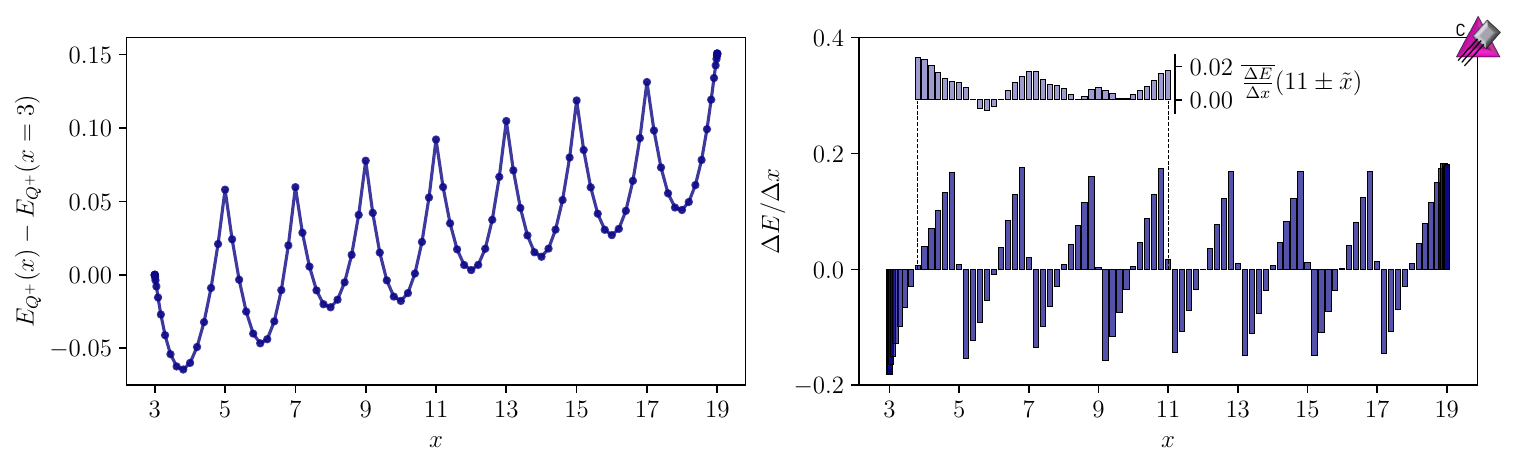}
\caption{
The energy as a function of the position of the heavy-$Q^+$ 
with the initial energy subtracted (left) and rate of energy loss $\Delta E/ \Delta x$, as defined in Eq.~\eqref{eq:dedxFD} (right).
The inset on the right panel shows the symmetrized average of $\Delta E/ \Delta x$ around $x=11$.
This simulation was performed with $L=12$, $m=0.1$, and $g=0.8$.
The heavy-$Q^+$ 
is initially at staggered site $x_0=3$ and moves to $x_f=19$, using a velocity profile with $v_{\rm max}=0.2$.
A purple icon in the upper right indicates that this calculation was performed using a classical computer~\cite{Klco:2019xro}.}
\label{fig:dEdxVAC}
\end{figure}
Figure~\ref{fig:dEdxVAC} shows the change in the total energy,
$E_{Q^+}(x)$
with parameters $m=0.1$ and $g=0.8$, using the same classical trajectory as in Fig.~\ref{fig:TRAJv0p2a0p2}, 
but with $L=12$ and $x_f = 19$.
Also shown is the instantaneous energy-loss $\Delta E/\Delta x$ 
defined by a 
finite-difference approximation to the energy loss 
at position $x$ at time $t$ during a Trotter step of size $\Delta t$,
\begin{equation}
    \frac{dE}{dx}(x) \ \rightarrow \  \frac{\Delta E}{\Delta x} \ = \ 
\frac{E_{Q^+}(t+\Delta t) - E_{Q^+}(t-\Delta t)}{x(t+\Delta t) - x(t-\Delta t)}
    \ .
\label{eq:dedxFD}
\end{equation}
The saw-tooth structure of the energy is due to the staggering of charges.
Energy decreases as the heavy-$Q^+$ moves toward an even-numbered (electron) site and away from an odd-numbered (positron) site, due to the Coulomb interaction.
Similarly, the energy increases as the heavy-$Q^+$ moves away from an electron site and toward a positron site.
This structure likely results from our implementation of motion across the lattice, 
and could be mitigated by smoothing the charge evolution over more than two lattice sites,
and by  decreasing the lattice spacing.
The inset of the right panel shows the sum of the contributions symmetrized around the midpoint of the lattice ($x=11$), $\overline{\frac{\Delta E}{\Delta x}}(11\pm \tilde{x})=\frac{1}{2}[\frac{\Delta E}{\Delta x} (11+\tilde{x})+\frac{\Delta E}{\Delta x} (11-\tilde{x})]$, with $0\leq\tilde{x}\leq 7$.
The strong cancellations between the positive and negative contributions indicates that the net 
$\Delta E / \Delta x$ is small when averaged across  lattice sites, 
much smaller than the magnitude of typical instantaneous values,
but importantly not equal to zero.  
This demonstrates that there is a net energy 
gain per unit length in the light degrees of freedom
as the heavy charge moves across the lattice 
vacuum, corresponding
to the production of light hadrons, and a net energy loss of heavy charge.

Lorentz breaking operators in the Hamiltonian
will, in general, contribute terms that are suppressed by powers of the lattice spacing, 
${\cal O}(a^n)$, with the lowest contribution at
${\cal O}(a^2)$ for the KS Hamiltonian~\cite{Kogut:1974ag,Banks:1975gq}.
Matrix elements of the Lorentz-breaking operators with a moving 
heavy-$Q^+$ are expected to give rise to contributions to observables that scale as 
${\cal O}\left(a^2 v^2\right)$ 
at low velocities
after 
parity considerations and renormalization of the Lorentz-preserving contributions.
As the velocity of the heavy charge approaches the speed of light, $v\rightarrow 1$, 
higher order terms will become increasingly important. 
\begin{figure}[!t]
\centering
\includegraphics[width=\textwidth]{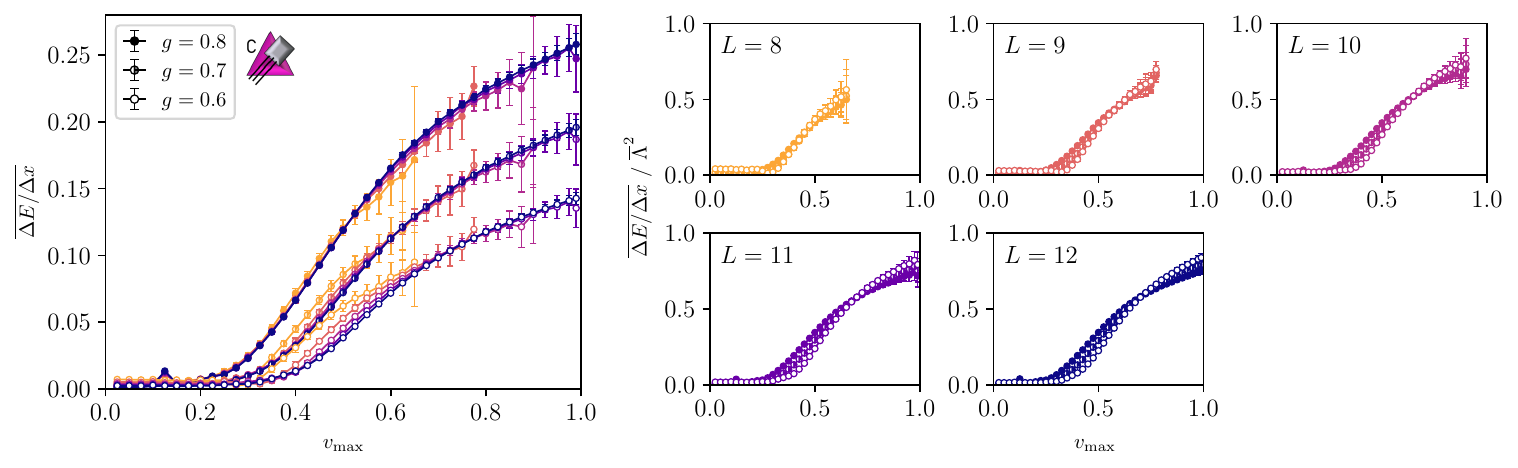}
\caption{
Left:
the lattice-averaged rate of energy loss $\overline{\Delta E/\Delta x}$, for a heavy-hadron
moving from $x_0=3$ to $x_f=2(L-3)+1$ across the vacuum as a function of velocity.
Right: the same as the left panel
but rescaled by the square of the heavy-hadron mass $\overline{\Lambda}^2$.
Results are shown for $L=\{8,\ldots,12\}$ and $g=\{0.8,0.7,0.6\}$ with $m/g=0.125$. 
Decreasing $g$ with $m/g$ fixed effectively decreases the lattice spacing.
}
\label{fig:AVdEdxVsvelocity}
\end{figure}
Figure~\ref{fig:AVdEdxVsvelocity} shows a 
lattice-averaged rate of energy loss, 
$\overline{\Delta E/\Delta x}$, as a function of the velocity of the heavy-$Q^+$.
$\overline{\Delta E/\Delta x}$ is determined by a linear fit to $E_{Q^+}(x)$ 
in the region of constant velocity, 
defined by $v \geq (v_{\rm max}-0.01)$.
Using our trajectories, there is an upper bound on $v_{\text{max}}$ 
for a given simulation volume, leading to incomplete curves for the smaller $L$ in Fig.~\ref{fig:AVdEdxVsvelocity}.
The results are consistent with
expected quadratic dependence on $v$ for low velocity.
The lattice-spacing dependence can also be probed by decreasing the coupling $g$ while keeping $m/g$ fixed (effectively decreasing $a$).
The results in Fig.~\ref{fig:AVdEdxVsvelocity} verify that the energy loss decreases as the lattice spacing decreases. 
This can be made more manifest by forming a dimensionless quantity between two physical quantities, that vanishes in the continuum.
The right panels of Fig.~\ref{fig:AVdEdxVsvelocity}
show $\overline{\Delta E/\Delta x}$ rescaled by the square of the heavy hadron mass, $\overline{\Lambda}^2$.
Keeping the physical heavy hadron mass fixed gives $\overline{\Lambda}\sim a$, 
whereas $\overline{\Delta E/\Delta x}$ is expected to scale at least as $\sim a^3$.
The rescaled energy loss is indeed seen to decrease for smaller lattice spacing up to a $v_{\text{max}} \sim 0.7$,  
where this analysis appears to break down.

\begin{figure}[!t]
\centering
\includegraphics[width=\textwidth]{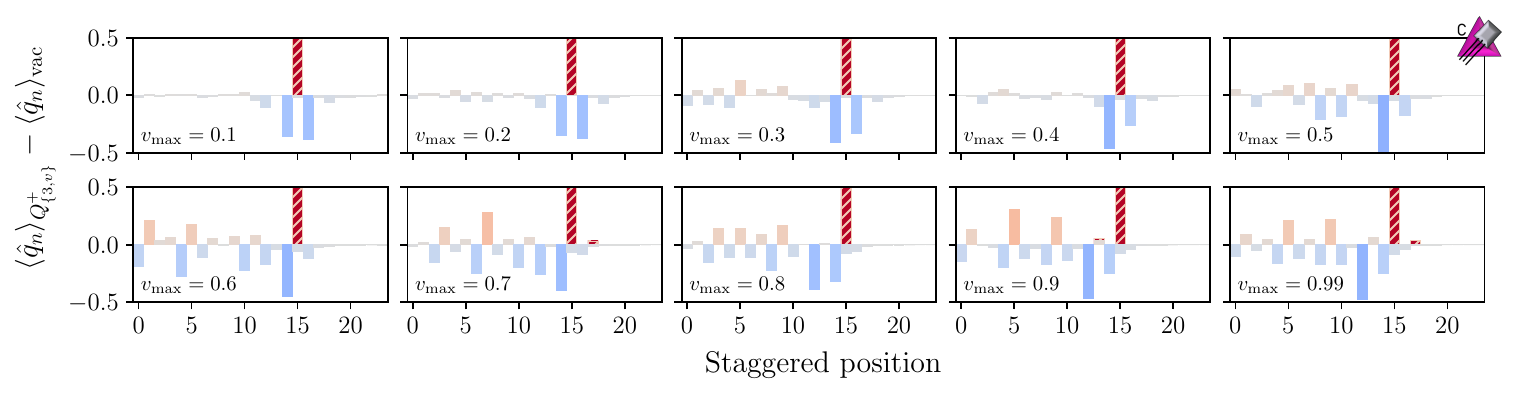}
\caption{
The vacuum subtracted charge density in the light degrees of freedom (solid bars) 
as a function of velocity, 
when the heavy-$Q^+$ (dashed bar) is at (or near) $x=15$.
These simulations were performed with $L=12$, $m=0.1$, and $g=0.8$.
}
\label{fig:QplusVac}
\end{figure}

As the heavy charge moves through the vacuum, the light charges re-arrange to dynamically screen the $Q^+$.
The charge density in the Lorentz-invariant continuum remains localized around the heavy charge, with a symmetric profile that is increasingly Lorentz-contracted with increasing velocity.
On the lattice, this picture changes due to Lorentz symmetry being broken, and particles having a modified dispersion relation. 
Importantly, the light charges have a maximum velocity $v_{\star}$ 
(see Eq.~\eqref{eq:Vstar} for $g=0$) that is less than the speed of light.
These lattice effects are illustrated in Fig.~\ref{fig:QplusVac}, which shows the charge density when the heavy-charge is at $x=15$ for a range of velocities.\footnote{This is not an ideal 
observable because the charge density in the wake of the heavy charge will fluctuate with time.
This is why the charge density behind the heavy charge is unusually small for $v_{\text{max}}=0.4$ for that specific time (compared to $v_{\text{max}}=0.3$ or $v_{\text{max}}=0.5$).}
At $t=0$ in our simulations, there is a symmetric distribution of charges screening the $Q^+$ 
(up to boundary effects).
For $v_{\text{max}} \lesssim 0.3$, 
this screening cloud largely
travels with the heavy charge, reproducing continuum expectations.
However, as $v_{\text{max}}$ becomes comparable to $v_{\star}$, the light charges cannot keep up with the heavy charge.
The light charges are dragged behind the heavy charge, and the profile becomes more asymmetric with increasing velocity.
This asymmetric charge distribution exposes the vacuum to a strong electric field, causing particle (hadron) production in the wake of the moving charge.
This is seen in the fluctuations in the light degrees of freedom on the opposite (left) side of the lattice, where light hadrons have a non-zero probability of being produced during the motion.

\begin{figure}[!t]
\centering
\includegraphics[width=\textwidth]{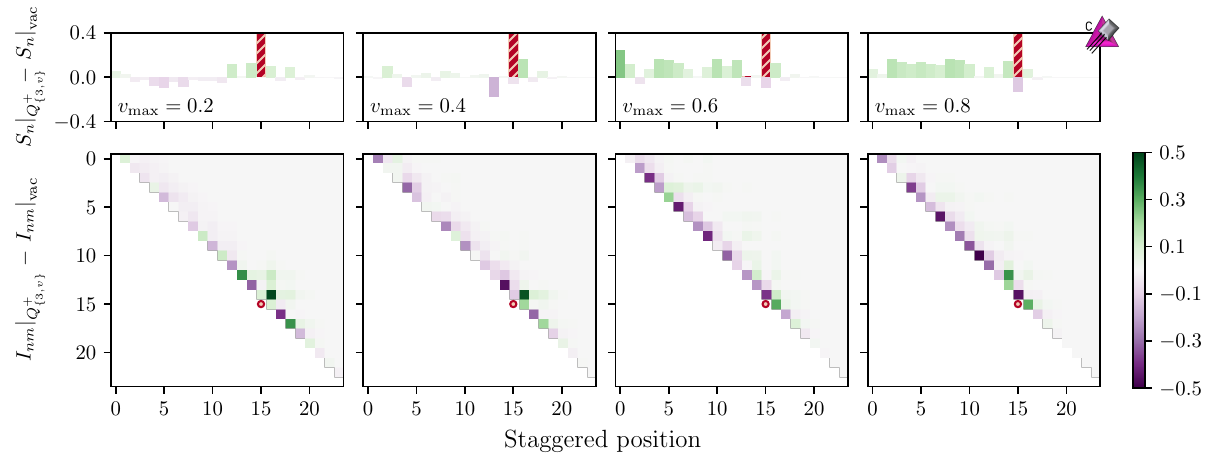}
\caption{
The vacuum subtracted single-site entanglement entropy in the light degrees of freedom (top) and mutual information (bottom) when the moving heavy-$Q^+$ is at $x=15$ (dashed red bar). 
Being a symmetric matrix, only the upper triangular part of the mutual information is shown.
The simulation parameters are the same as those used to generate Fig.~\ref{fig:QplusVac}.
}
\label{fig:QplusSiInm}
\end{figure}

The role of quantum correlations in strongly interacting systems is an area of active research, with pioneering work connecting entanglement to the confinement and chiral phase transitions in QCD~\cite{Beane:2018oxh,Beane:2019loz,Beane:2020wjl,Beane:2021zvo,Liu:2022grf,Miller:2023ujx,Liu:2023bnr}, with parallel works on low-energy nuclear systems~\cite{Robin:2020aeh,Bulgac:2022cjg,Johnson:2022mzk,Gu:2023aoc,Hengstenberg:2023ryt,Perez-Obiol:2023wdz}, high-energy processes~\cite{Kharzeev:2017qzs,Cervera-Lierta:2017tdt,Baker:2017wtt,Kharzeev:2021yyf,Gong:2021bcp,Florio:2024aix}, and quantum field theories~\cite{Srednicki:1993im,Marcovitch:2008sxc,Klco:2020rga,Klco:2021biu,Klco:2021cxq,Klco:2023ojt,Parez:2023uxu,Florio:2023mzk}.
In the continuum, and similar to the charge density,
it is expected that disturbances in the entanglement above the vacuum will be localized around the position of the heavy charge.
However, on the lattice, the production of hadrons in the wake of the $Q^+$ alter the localized entanglement signatures.
The single-site entanglement entropy $S_n=- {\rm Tr}(\rho_n \log_2 \rho_n)$ 
is related to the purity of the reduced density matrix, $\rho_n$, 
on site $n$, 
and is shown in Fig.~\ref{fig:QplusSiInm} 
when the heavy-$Q^+$ is at $x=15$ for a selection of $v_{\text{max}}$ (same situation as in Fig.~\ref{fig:QplusVac}).
For small
$v_{\text{max}}$, the continuum expectation of entanglement entropy localized around the heavy-$Q^+$ is recovered.
For larger $v_{\text{max}}$, 
considerable entanglement entropy is generated in the wake of the moving charge, consistent with a lattice artifact that scales as ${\mathcal O}(a^2 v^2)$ at low velocities.
To investigate correlations between sites, the bottom panels of Fig.~\ref{fig:QplusSiInm} 
show the mutual information $I_{nm}$.
The mutual information between sites $n$ and $m$ is defined as 
$I_{nm}=S_n+S_m-S_{nm}$, where $S_{nm} = - {\rm Tr}(\rho_{nm} \log_2 \rho_{nm})$ 
is the entanglement entropy of the two-site reduced density matrix.
This quantity shows that the correlations produced by the moving charge are 
short range, with a scale naturally set by confinement.
For this particular system, these entanglement measures are providing a qualitative picture that is consistent with the charge densities in Fig.~\ref{fig:QplusVac}.\footnote{The difference in sign between $I_{nm}$ and $S_n$ in Fig.~\ref{fig:QplusSiInm} (especially in the last column) is due to $S_{nm}$ being much larger with the moving charge than in the vacuum.}

\begin{figure}[!t]
\centering
\includegraphics[width=\textwidth]{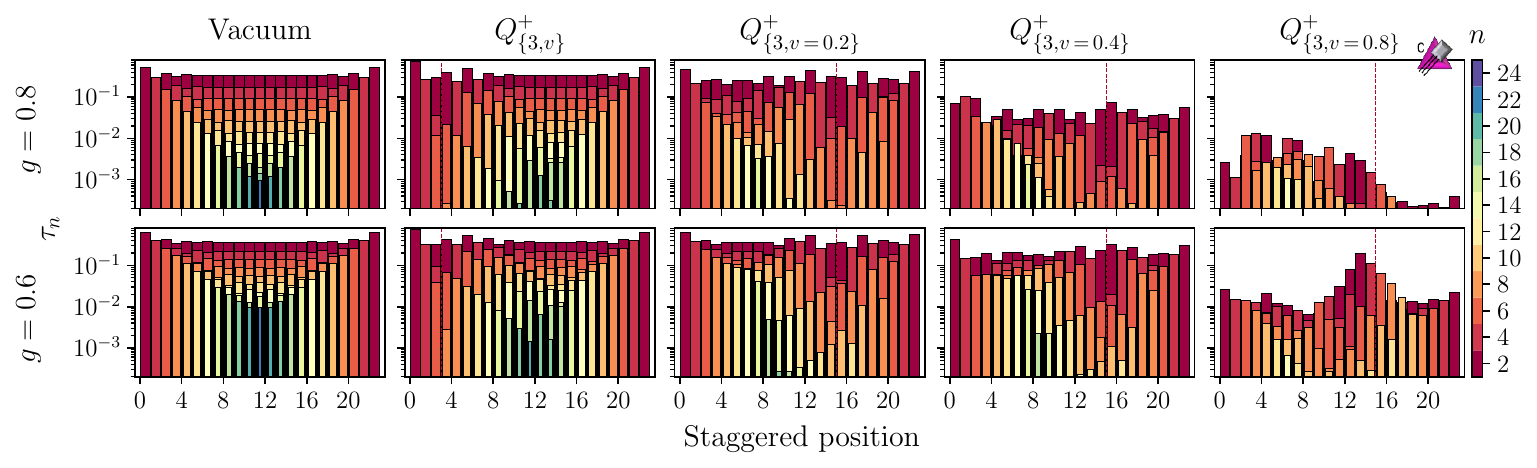}
\caption{
The $n$-tangles for $n=\{2,\ldots,24\}$
associated with the vacuum state (first column), with a moving heavy-$Q^+$ at its starting position $x_0=3$ (second column), and near position $x=15$ (on its way to $x_f = 19$) with velocity $v_{\rm max}=0.2$ (third column), $v_{\rm max}=0.4$ (fourth column), and $v_{\rm max}=0.8$ (fifth column). 
The simulations were performed using $g=0.8$ (top row) and $g=0.6$ (bottom row) with $m/g=0.125$ and $L=12$.
The bars representing $\tau_n(|\psi\rangle)^{i_1,... i_n}$ are centered around the mid $\{i_1,\ldots,i_n\}$ point, with $i_k=i_1+(k-1)$.}
\label{fig:ntangs}
\end{figure}

Quantum correlations beyond two sites can be characterized by
the $n$-tangle~\cite{Wong:2001} of a pure state $\vert \psi \rangle$, defined by
\begin{equation}
\tau_n(|\psi\rangle)^{i_1,... i_n} \ = \
|  \langle \psi |\tilde \psi \rangle  |^2
\ ,\ \ 
|\tilde \psi \rangle \ = \ \hat Y_{i_1}\hat Y_{i_2}\cdots\hat Y_{i_n}
\ |\psi\rangle^*
\ .
\label{eq:ntang}
\end{equation}
The $n$-tangle for odd-$n$ is only defined for $n=3$, and 
vanishes in this system by charge conservation.
Confinement suggests that the $n$-tangles should
fall off exponentially when the number of contributing lattice sites exceeds the confinement length. 
This can be seen in the first column of Fig.~\ref{fig:ntangs}, 
where the non-zero $n$-tangles of the vacuum state $| \psi_{\rm vac}\rangle$ are shown.
The second column of Fig.~\ref{fig:ntangs} shows the $n$-tangles with a heavy-$Q^+$ at its initial position $x_0=3$.
They deviate from those in the vacuum in a local region around the heavy charge, as expected for a system with a finite correlation length.
If there were no lattice discretization effects, deviations from the vacuum $n$-tangles would remain localized around the moving charge.
The lattice effects are illustrated in columns three through five, which show the $n$-tangles when the heavy-charge is at $x=15$ for a selection of velocities.
Relative to the initial state (column two), a moving charge modifies the $n$-tangles across the whole lattice. 
These effects are magnified for larger velocities or coarser lattice spacings, compare $g=0.8$ (top row) and $g=0.6$ (bottom row), as expected for a lattice artifact.
However, these lattice artifacts behave noticeably different than the local observables examined in the previous paragraphs.
The $n$-tangles are more fragile than other observables; with $g=0.8$, the $2$-tangle is diminished by more than $10\times$ for a velocity of $v=0.4$.\footnote{Note that the finite time step in our time evolution, 
$\Delta t$ in Eq.~\eqref{eq:tevol}, 
induces errors on the order of $\sim 10^{-3}$ in the $n$-tangles. 
While decreasing the value of $\Delta t$ modifies the values of the $n$-tangles, it does not change the qualitative features observed.
This is much more severe than, for example, the deviation of the charge density shown in Fig.~\ref{fig:QplusVac}.
In addition, while the deviations in the charge density are restricted to the region behind the moving charge, the $n$-tangle is significantly destroyed across the entire lattice.
These differences are likely because the $n$-tangle is not a local observable 
due to the complex conjugation in Eq.~\eqref{eq:ntang}.}

The suppression of the $n$-tangles is a striking result, compared to the single-site 
and two-site entropy.
A possible explanation for this difference is that the $n$-tangles are not capturing all
of the entanglement in the system 
(e.g., when evaluated in the GHZ and W states, $S_n$ and $I_{mn}$ are non-zero while certain $n$-tangles are zero~\cite{Wong:2001,Illa:2022zgu}). 
The results from these entanglement measures point to 
observables of the system evolving toward those of a classically mixed ensemble,
as predicted for pure state evolution consistent with the 
{\it Eigenstate Thermalization Hypothesis}
(ETH)~\cite{Srednicki:1994mfb} 
(a similar connection was found in Ref.~\cite{Florio:2024aix}).
To further understand how the entanglement structure evolves, other measures, such as the negativity~\cite{Zyczkowski:1998yd,Vidal:2002zz} or non-stabilizerness entanglement (magic)~\cite{Bravyi:2012,Leone:2021rzd,Tirrito:2023fnw}, could be studied.
We remind the reader that velocity dependence (aside from Lorentz contraction) of the observables calculated in this section are lattice artifacts that will vanish in the continuum.

\subsection{A Heavy-\texorpdfstring{$Q^+$}{} Moving Through a Dense Medium}
\label{sec:HQmed}
\noindent
A main objective of this work is to develop machinery for quantum simulations of dynamics in dense matter.
The previous subsection quantified the energy loss and other lattice artifacts that are already present for a heavy charge moving across the vacuum.
This provides a benchmark to compare with the results of in-medium simulations.
Matter is introduced into our simulations by including one or more static heavy-$Q$s, 
whose positions are fixed
in time.
For well separated static charges, the ground state consists of a grid of heavy hadrons at rest.
For tightly packed static charges, the screening clouds merge together, analogous to the electron sea in a metal.
The parameters used in this work give rise to screening 
with high fermion occupation numbers localized over a couple of staggered sites.
Because of this, Pauli blocking plays a significant role in the dynamics.
Combined with the kinematic restrictions of one dimension, 
evolution within the medium leads
to interesting phenomena, 
such as significant distortions to the screening profiles that is 
more pronounced in the leading edge of the collision.
In the continuum, collisions between hadrons 
are inelastic above a given
threshold invariant mass, depending on the hadronic spectrum.
In our simulations, with the hadron velocity fixed to $v_{\text{max}}$ throughout 
the collision, 
hadron production is possible for all kinematics.
The simulations in this section are all performed
with $L=12$ and a relatively low heavy-$Q^+$ velocity of $v_{\text{max}}=0.2$ to minimize lattice artifacts.
The rest of the parameters defining the classical trajectory of the heavy-$Q^+$ are the same as in the vacuum simulations of the previous section.

\subsubsection{A Heavy-\texorpdfstring{$Q^+$}{} Incident upon One Static-\texorpdfstring{$Q^+$}{} }
\label{sec:QonQ}
\noindent
The simplest system to begin studies of energy loss in matter is that of 
a neutralized heavy-$Q^+$ moving past another neutralized heavy-$Q^+$ that is fixed in place.
Initially, we prepare the ground state of the system in the presence of two heavy-$Q^+$s: the moving charge at $x_0=3$ and the static charge at $x=11$. 
The total charge of the light 
sector in the
ground state is $q_{\text{tot}}=-2$.
This initial wavefunction is time-evolved 
using Eq.~\eqref{eq:tevol}, and the energy loss and charge density are compared to the results from the vacuum simulations in the previous section.

\begin{figure}[!t]
\centering
\includegraphics[width=\textwidth]{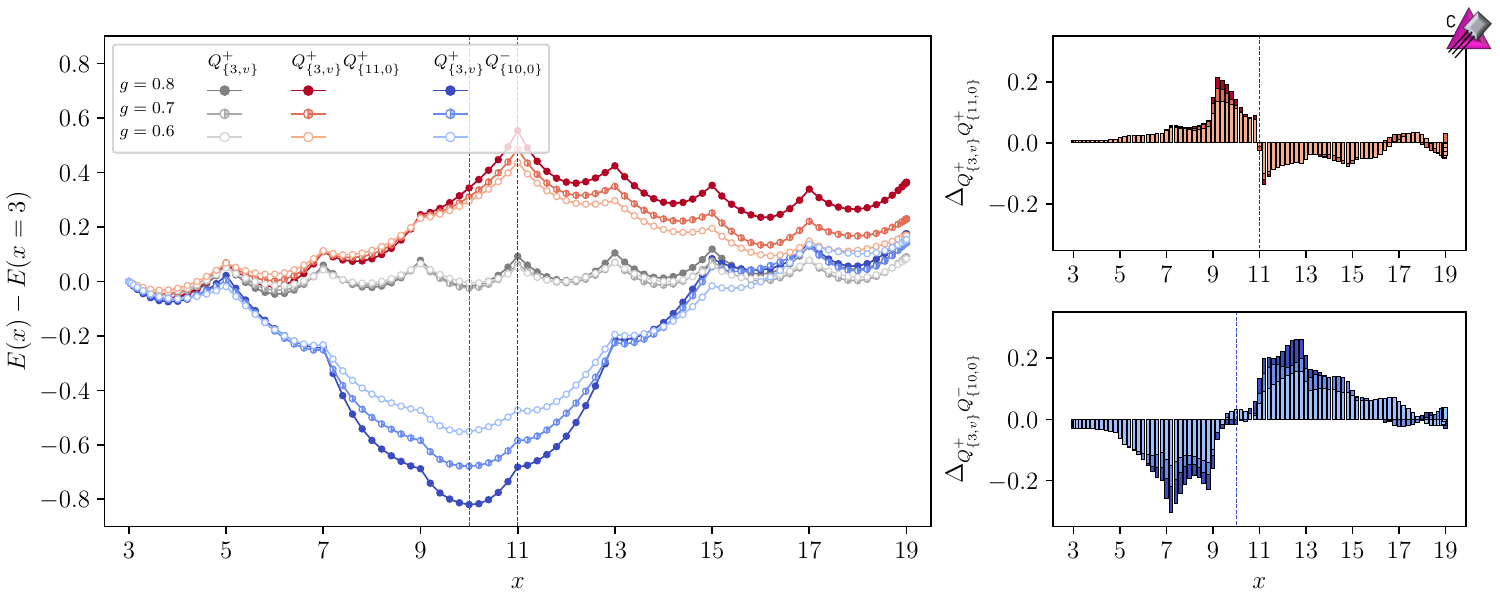}
\caption{
The energy as a function of the position  of the heavy-$Q^+$ with the initial energy subtracted (left) and rate of energy loss, as defined in Eq.~\eqref{eq:DeltaMed} (right).
The heavy-$Q^+$ velocity is $v_{\rm max}=0.2$ and the static heavy-$Q^+$ ($Q^-$) is located at $x=11$ ($x=10$).
These simulations were performed with $L=12$ for  $g=\{0.8,0.7,0.6\}$ and $m/g=0.125$.
The curve labeled as $Q^+_{\{3,v\}}$ denotes a heavy-$Q^+$ initially at $x_0=3$ moving though the vacuum.
The curve denoted by $Q^+_{\{3,v\}} Q^+_{\{10,0\}} $ denotes a heavy-$Q^+$ initially at $x_0=3$ moving past a static heavy-$Q^+$ at $x=10$, and similarly for the curve  $Q^+_{\{3,v\}} Q^-_{\{11,0\}}$.
The dashed red (blue) vertical lines mark the position of the positive (negative) static charge.
}
\label{fig:QplusQplusminus}
\end{figure}

\begin{figure}[!t]
\centering
\includegraphics[width=\textwidth]{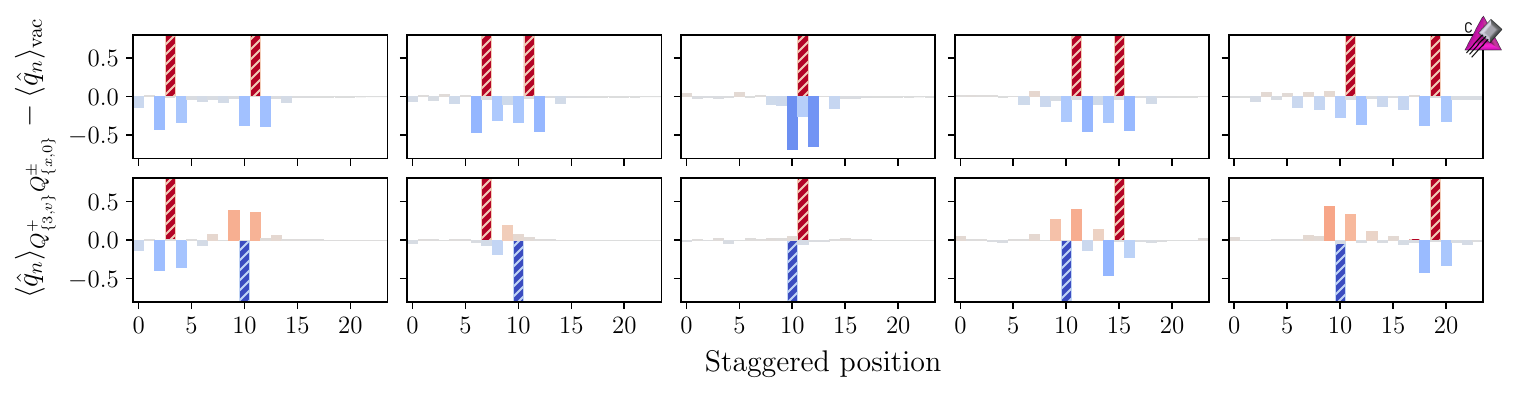}
\caption{
The vacuum subtracted charge distribution 
in the light degrees of freedom (solid bars) shown for when the moving heavy-$Q^+$ is at $x=\{3,7,11,15,19\}$ (left-to-right).
The results in the upper (bottom) panels are for a static heavy-$Q^+$ ($Q^-$) located at $x=11$ ($x=10$).
The dashed bars denote the location of the heavy-charges, and simulation parameters $v_{\text{max}}=0.2$, $m=0.1$, $g=0.8$, and $L=12$ have been used.
}
\label{fig:QplusQplusQminus_charge}
\end{figure}

Figure~\ref{fig:QplusQplusminus} shows the energy loss as a function of the 
position of the moving heavy-$Q^+$ in the presence of the static heavy-$Q^+$ (red points).
To remove some of the lattice artifacts, it is useful to define the vacuum-subtracted quantity $\Delta_{Q^+_{\{3,v\}}Q^\pm_{\{x',0\}}}$,
\begin{equation}
    \Delta_{Q^+_{\{3,v\}}Q^\pm_{\{x',0\}}} \ = \ \left. \frac{\Delta E}{\Delta x} \right|_{Q^+_{\{3,v\}} Q^\pm_{\{x',0\}}}
    - \left. \frac{\Delta E}{\Delta x} \right|_{Q^+_{\{3,v\}}}
    \ ,
    \label{eq:DeltaMed}
\end{equation}
where $x'$ is the staggered site of the static charge.
As expected, the energy loss receives its largest contributions when the screening 
of the two heavy-charges overlap.
The change in energy is more rapid
on the leading edge of the collision than the trailing edge.
This suggests that the initial collision is similar to a violent quench whereas the trailing interactions are occurring 
closer to equilibrium.
Further insight into the mechanisms involved in this process can be gathered from Fig.~\ref{fig:QplusQplusQminus_charge}, which shows the evolution of the charge density.
Particularly striking is the charge density around the static charge after the collision (top-right panel).
Compared to the 
results in vacuum, Fig.~\ref{fig:QplusVac}, the charge density in-medium is more de-localized.
This is indicative of excitations of the static hadron, and of light hadron production in the collision.
Note that the difference in the charge distributions of the static and dynamic charge at $t=0$ (left panel) are due to boundary effects.
%

\subsubsection{A Heavy-\texorpdfstring{$Q^+$}{} Incident upon One Static-\texorpdfstring{$Q^-$}{}}
\label{sec:QonQbar}
\noindent
By placing a static $Q^-$ in the volume instead of a static $Q^+$, 
collisions of a heavy-meson with a heavy-anti-meson can  also be studied.
One difference between the simulations involving two equal charges is that the effects of Pauli blocking are less significant; the electrons surrounding the moving heavy-$Q^+$ are not Pauli blocked by the positrons surrounding the static $Q^-$.
In addition, there is now a particle-antiparticle annihilation channel open during the collision.
The energies during these simulations are shown in Fig.~\ref{fig:QplusQplusminus} (blue points).
As expected, they are seen to be lower for the oppositely charged heavy-$Q$s than for the same
charged heavy-$Q$s.
The right panels show that the energy changes much more rapidly than for the two colliding $Q^+$s, due to the lack of Pauli blocking.
Also, the net change in energy in this process is noticeably less than 
when a  heavy-$Q^+$ passes a static heavy-$Q^+$.

The charge density is shown in the lower panels of Fig.~\ref{fig:QplusQplusQminus_charge}.
For this relatively low velocity,
the screening vanishes when the charges are on top of each other (middle column), because the total net 
heavy charge is zero.
It is interesting to look 
at the charge distribution surrounding the static-$Q^-$ after the moving charge has passed.
When the positive charge is at $x=15$ (fourth column) the charge distribution surrounding the $Q^-$ has a dipole moment pointing to the right.
However, when the positive charge is at $x=19$ (fifth column), the dipole moment is 
pointing to the left.
This suggests that the heavy-$Q^-$ hadron  is left 
in an excited state characterized by a time-dependent dipole moment.
Such excitations have recently been identified in dynamical simulations of nuclei moving through dense neutron matter~\cite{Pecak:2024zqq}.

\subsubsection{A Heavy-\texorpdfstring{$Q^+$}{} Incident upon Two Static-\texorpdfstring{$Q^+$}{}s}
\label{sec:QonQQ}
\noindent
A medium of multiple neutralized static $Q^+$s allows for an exploration of in-medium quantum coherence, beyond those involved in heavy-hadron  collision.
Limited by the sizes of lattice volumes available for classical simulation, we consider two static $Q^+$s located in the middle of the lattice and separated by one or two spatial sites.
Quantum correlations between tightly packed static charges are expected to have a large effect on the energy loss, which will not simply be an incoherent sum of the energy lost to each static charge separately.
\begin{figure}[!t]
\centering
\includegraphics[width=\textwidth]{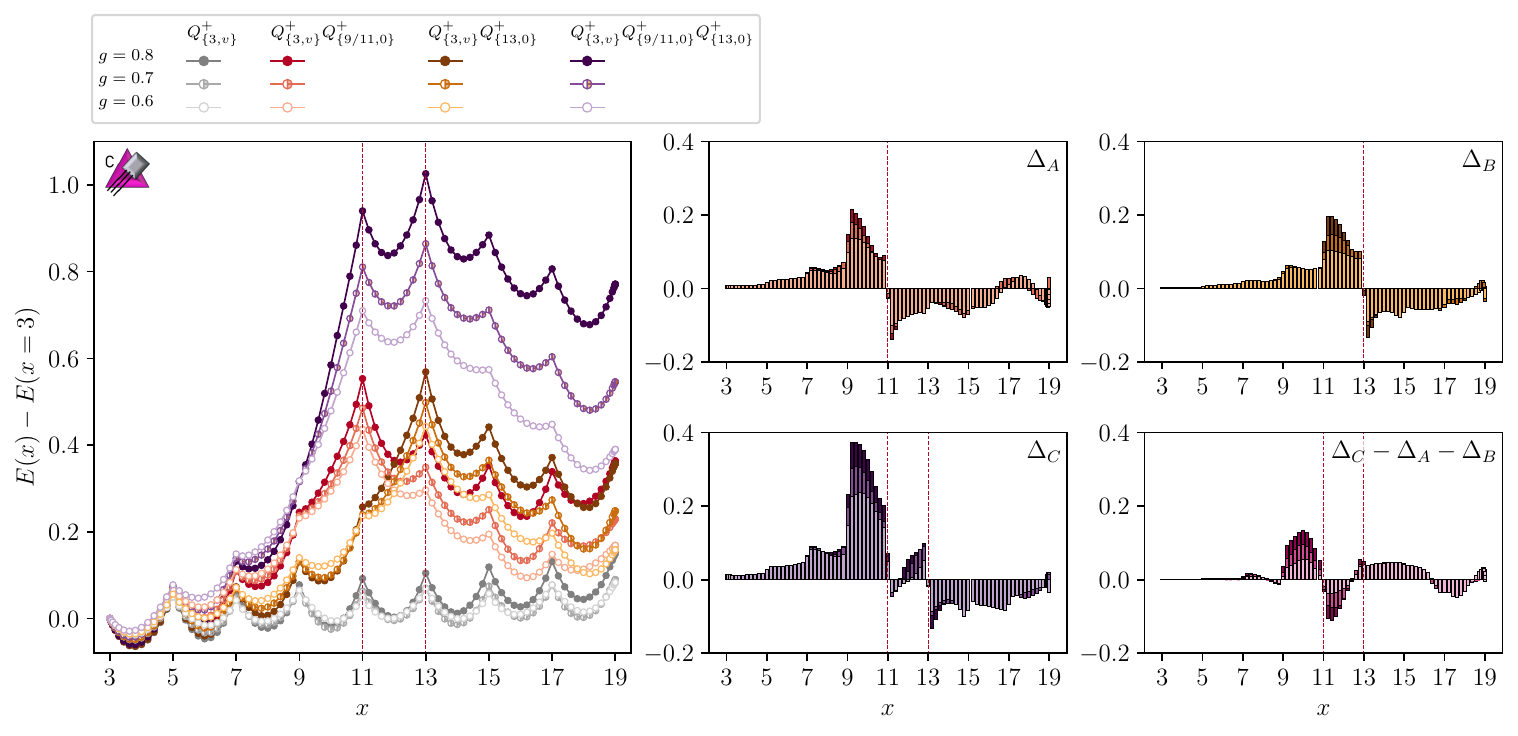}
\includegraphics[width=\textwidth]{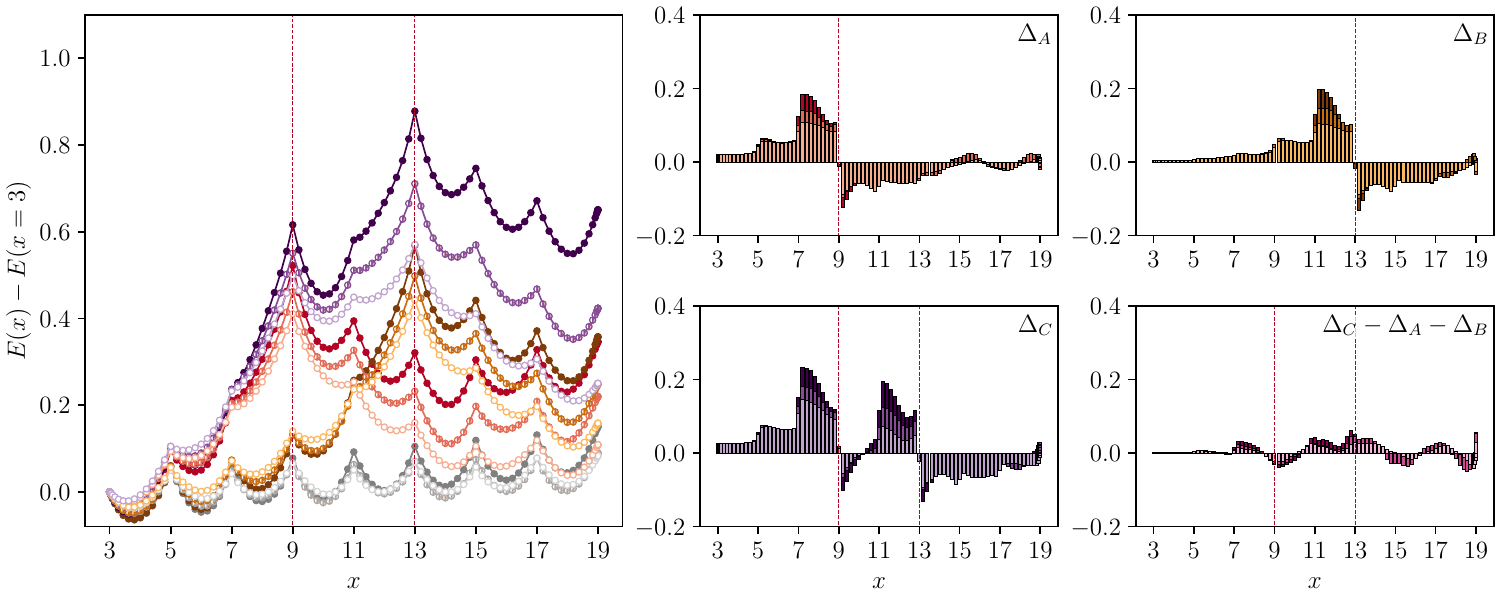}
\caption{
The energies as a function of position of the heavy-$Q^+$ moving 
through two static heavy-$Q^+$s located at $x=\{11,13\}$ (top) and $x=\{9,13\}$ (bottom)
after the initial-state energy has been subtracted.
The left panel shows the total energy, $E(x)$, for the 
$Q^+_v$-$Q^+Q^+$ systems (purple), 
the individual $Q^+_v$-$Q^+$ systems (red and orange), 
and the $Q^+_v$-vacuum system (gray).
The right panels show $\frac{\Delta E}{\Delta x}$ for these same systems
as defined in  Eq.~\eqref{eq:Deltadef}.
The simulations were  performed with $L=12$ for different values of $g=\{0.8,0.7,0.6\}$ and $m/g=0.125$.
The heavy-$Q^+$ started at staggered site $x=3$ and moved to $x=19$, using a velocity profile 
with $v_{\text{max}}=0.2$.
}
\label{fig:QplusONQplusQplus}
\end{figure}
Figure~\ref{fig:QplusONQplusQplus} shows the energy loss as a function of position of the moving heavy-$Q^+$ in the presence of two static-$Q^+$ 
(for separations of one (upper) and two (lower) spatial sites).
To isolate the effects of in-medium quantum coherence, 
it is useful to define the following quantities,
\begin{align}
    \Delta_A & \ = \ \left. \frac{\Delta E}{\Delta x} \right|_{Q^+_{\{3,v\}} Q^+_{\{x',0\}}}
    - \  \left. \frac{\Delta E}{\Delta x} \right|_{Q^+_{\{3,v\}}}
     , \quad \quad \Delta_B \ = \ \left. \frac{\Delta E}{\Delta x} \right|_{Q^+_{\{3,v\}} Q^+_{\{x'',0\}}}
    - \  \left. \frac{\Delta E}{\Delta x} \right|_{Q^+_{\{3,v\}}}  ,
    \nonumber\\
    \Delta_C & \ = \ \left. \frac{\Delta E}{\Delta x} \right|_{Q^+_{\{3,v\}} Q^+_{\{x',0\}} Q^+_{\{x'',0\}}}
    - \  \left. \frac{\Delta E}{\Delta x} \right|_{Q^+_{\{3,v\}}}
     ,
    \label{eq:Deltadef}
\end{align}
where $x', x''$ indicate the lattice sites of the 
static-$Q^+$s, 
and these quantities are shown in the right panels of Fig.~\ref{fig:QplusONQplusQplus}.
The combination $\Delta_C - \Delta_A - \Delta_B$ is a measure of in-medium quantum coherence, and is seen to be more significant for static-$Q^+$s separated by one spatial site compared to two.
Even at low velocities, the energy loss function is sensitive to the increased fermion occupancy and quantum coherence present in dense systems.
As in the case of a single static heavy hadron,
there is a clear asymmetry between the interactions of the leading light degrees of freedom and the trailing ones.


\section{Quantum Simulations}
\label{sec:Qsims}
\noindent
The initial state for the simulations performed in this work is the ground state 
in the presence of   background charges, $\vert \psi_{\text{vac}} \rangle_{Q_{\{x\}}}$. 
Without these background  charges, the total charge of the vacuum is $q_{\text{tot}} = 0$.
In the presence of the charges, the ground state of the system re-arranges in such a way that 
$q_{\text{tot}} +  Q_{\text{tot}} = 0$  for sufficiently large lattices.\footnote{
For a finite lattice size, 
there is a regime of large $m/g$ for which $q_{\text{tot}} = 0$ 
when $Q_{\rm tot}\neq 0$, and the ground state of the system is charged.
For the relatively small $m/g=0.125$ used in this work, the ground state has 
$q_{\text{tot}} +  Q_{\text{tot}} = 0$.}
In this phase, the charges are completely screened over the scale of a confinement length $\xi \sim m_{\text{hadron}}^{-1}$.
Outside of this screening length, the system is locally in the vacuum {\it without} static charges, $\vert \psi_{\text{vac}} \rangle$.
These observations inform an efficient and scalable method for preparing ground states  
in the presence of background  charges on a quantum computer.
A key ingredient in the method for state preparation is the Scalable-Circuit-ADAPT-VQE (SC-ADAPT-VQE) algorithm that will be briefly reviewed in the next subsection.

\subsection{SC-ADAPT-VQE}
\label{sec:SCADAPT}
\noindent
SC-ADAPT-VQE is an algorithm for efficiently preparing states with localized correlations on a quantum computer~\cite{Farrell:2023fgd,Farrell:2024fit}. It utilizes symmetries and hierarchies in length scales to determine systematically improvable quantum circuits for preparing a desired state that can be scaled to arbitrarily large system sizes. 
It is a variational algorithm where parameterized quantum circuits 
are optimized to prepare a desired state.
Crucial to the SC-ADAPT-VQE workflow is that circuit optimization can be performed using
a series of small system sizes accessible to 
{\it classical} computers (or a small partition of an
quantum computer). 
The variational parameters specifying the circuits can then be extrapolated to 
prepare the desired state on a large system using a {\it quantum} computer.
This technique was used to prepare the Schwinger model vacuum~\cite{Farrell:2023fgd} and a hadron wavepacket~\cite{Farrell:2024fit} on up to 112 qubits of IBM's quantum computer.

The initial steps of SC-ADAPT-VQE parallel those of ADAPT-VQE~\cite{Grimsley_2019}, and are the following:
\begin{itemize}
    \item[1.] Define a pool of operators $\{ \hat{O} \}$ that respect the symmetries of the prepared state. 
    Scalability and phenomenological considerations are used to inform the content of the operator pool.  \item[2.] Initialize a state $\lvert \psi_{{\rm ansatz}}\rangle$ with the quantum numbers of the target state $\lvert \psi_{{\rm target}}\rangle$. 
    \item[3.] Determine a quantity that measures the quality of the ansatz state. 
    For ground states this can be the energy $E = \langle \psi_{\text{ansatz}}\vert \hat{H} \vert \psi_{\text{ansatz}}\rangle$.
    \item[4.] For each operator in the pool $\hat{O}_i$ estimate the gradient of the 
    energy of the evolved ansatz state, 
    $ \frac{\partial}{\partial\theta_i}\left. E \right|_{\theta_i=0} 
    = i\langle \psi_{\text{ansatz}}\vert [\hat{H},\hat{O}_i] \vert \psi_{\text{ansatz}}\rangle$. 
    This is one way of 
    estimating how $\hat{O}_i$ changes the energy.
    \item[5.] The operator $\hat{O}_n$ with the largest magnitude gradient is used to
    update the ansatz:  
    $\lvert \psi_{{\rm ansatz}} \rangle \to e^{i \theta_n \hat{O}_n}\lvert \psi_{{\rm ansatz}} \rangle$.
    \item[6.] Optimize the  parameters $\{\theta_i\}$ to minimize the energy.
    The previously optimized values for $\theta_{1,...,n-1}$ and $\theta_n=0$, are used as initial conditions.
    \item[7.] Return to step 4 until  a predetermined convergence threshold is met.
\end{itemize}
ADAPT-VQE returns an ordered sequence of unitary operators $\{ \hat{U}_i \} = \{ {\exp}(i \theta_i \hat{O}_i) \} $ that approximately prepares the target state.
In a quantum simulation, the $\{\hat{U}_i\}$ must be converted to a sequence of gates.
If this introduces, for example, Trotter errors, 
the unitary $e^{i \theta_n \hat{O}_n}$ in step 5 should be replaced by its Trotterized implementation.
SC-ADAPT-VQE has the additional steps:
\begin{itemize}
    \item[8.] Repeat steps 1-7 for a  range of lattice volumes $\{L_1, L_2, \ldots, L_N\}$ using a classical computer (or a small partition of a quantum computer).
    \item[9.] Extrapolate the parameters of the sequence of unitary operators 
    $\{\{\hat{U}_i\}_{L_1}, \{\hat{U}_i\}_{L_2}, \ldots, \{\hat{U}_i\}_{L_N}\}$ to the desired $L$ (which can be arbitrarily large).
    This sequence is expected to converge if the target state has localized correlations.
    \item[10.] Execute the circuit implementation of the sequence of extrapolated unitaries 
    $\{\hat{U}_i\}_L$ to prepare the desired state on a quantum computer.
\end{itemize}
The extrapolated sequence of circuits provides an explicit demonstration of localizable~\cite{Klco:2019yrb} and fixed-point~\cite{Klco:2020aud} quantum circuits.

\subsection{Preparing Ground States  with Background Charges}
\label{sec:StatVac}
\noindent
Consider preparing the ground state  
in the presence of a single positive background charge in the middle of the lattice $Q_{L-1}=+1$.
As argued above, the ground state 
has $q_{\text{tot}}=-1$, with the charge density localized around staggered site $L-1$.
The method that we will use to prepare $\vert \psi_{\text{vac}} \rangle_{Q^+_{L-1}}$ will have two steps:
\begin{itemize}
    \item[1.] Prepare a state $\vert \psi_{\text{init}}\rangle$ which has the qualitative features of $\vert \psi_{\text{vac}} \rangle_{Q^+_{L-1}}$ correct. 
     This state will be $\vert \psi_{\text{vac}} \rangle$ far from the background charge, and possess a local integrated charge of $q_{\text{tot}}=-1$ around the background  charge. 
     $\vert \psi_{\text{init}}\rangle$ is quantitatively correct everywhere except for a few correlation lengths around staggered site $L-1$.
    \item[2.] Modify the wavefunction around staggered site $L-1$. This builds the correct profile of the screening charges, and can be done with circuits that act locally around site $L-1$.
\end{itemize}

To construct $\vert \psi_{\text{init}}\rangle$, first initialize the strong coupling ground state 
in the presence of the background charge,
\begin{align}
\vert \Omega_0\rangle_{Q^+_{L-1}}  \ = \ \frac{1}{\sqrt{2}}\left (\hat{X}_{L-2}\vert \Omega_0\rangle \ + \ \hat{X}_{L}\vert \Omega_0\rangle \right ) \ ,
\end{align}
where $\vert \Omega_0\rangle = \vert 01\rangle^{\otimes L}$ is the strong-coupling vacuum without a background charge.
The $\hat{X}$ operators leads to electron occupation on the staggered sites next to the 
background charge.\footnote{
The state with an electron occupied on site $L-2$ and site $L$ are degenerate with the kinetic term turned off.
By time reversal symmetry, the state can be taken to be a real superposition, and it is found that the equal superposition with a $(+)$ has the lowest energy when the kinetic term is turned on.}
$\vert \Omega_0\rangle_{Q^+_{L-1}}$ has the desired property of charge $(-1)$ localized around the position of the background  charge.
Next, $\vert \psi_{\text{vac}}\rangle$ is prepared far away from the background  charge.
One way to accomplish this is to act with a unitary $\hat{U}^{\text{aVQE}}$, that prepares the vacuum when acting on the strong coupling vacuum, $\hat{U}^{\text{aVQE}}\, \vert \Omega_0 \rangle = \vert \psi_{\text{vac}} \rangle$. 
The problem of determining such a unitary with an efficient circuit implementation was recently addressed by the authors~\cite{Farrell:2023fgd}, 
and is an application of the SC-ADAPT-VQE algorithm outlined in Sec.~\ref{sec:SCADAPT}.
The use of SC-ADAPT-VQE 
to determine $\hat{U}^{\text{aVQE}}$ is reviewed in App.~\ref{app:SCADAPTvac}.
Acting this vacuum preparation unitary on $\vert \Omega_0\rangle_{Q^+_{L-1}}$,
\begin{align}
\vert \psi_{\text{init}}\rangle  \ = \ \hat{U}^{\text{aVQE}}\vert \Omega_0\rangle_{Q^+_{L-1}} \ ,
\label{eq:StatInit}
\end{align}
furnishes 
an initial state with the desired properties of having charge $(-1)$ localized around the 
background  charge and being $\vert\psi_{\text{vac}}\rangle$ away from the position of the 
background  charge.

For step 2, $\vert \psi_{\text{init}}\rangle$ is used as the initial state for another application of SC-ADAPT-VQE. 
The goal of this round of SC-ADAPT-VQE is to determine localized circuits that build the correct wavefunction in the region around the background  charge.
The target state $\vert \psi_{\text{vac}} \rangle_{Q^+_{L-1}}$ is the ground state of the Hamiltonian with a background charge $Q_{L-1}=+1$ and, since both the initial state and target state have $q_{\text{tot}}=-1$, the operators in the pool should conserve charge. 
In addition, as the Hamiltonian is real, the operators are also constrained by time-reversal invariance (operators with an odd number of $\hat{Y}$ in the Pauli string decomposition).
These constraints imply that there are no single-qubit operators, 
and a similar pool to 
that
used for preparing a hadron wavepacket in our previous work~\cite{Farrell:2024fit} is found to be effective,
\begin{align}
\{ \hat{O}  \}_{Q^+_{L-1}} \ &= \  \{ \hat{O}_{mh}(n,d)  \} \ , \nonumber \\
\hat{O}_{mh}(n,d) \ & \equiv \ \frac{i}{4}\left [ \hat{\sigma}^+_{L-1-n}\hat{Z}^{d-1}\hat{\sigma}^-_{L-1-n+d} \,  + \, {\rm h.c.}\ , \ \hat{Z}_{L-1-n}  \right ] \ = \ \frac{1}{2}\left (\hat{X}_{L-1-n} \hat{Z}^{d-1}\hat{Y}_{L-1-n+d}  -  \hat{Y}_{L-1-n} \hat{Z}^{d-1}\hat{X}_{L-1-n+d} \right ) ,
\label{eq:statPool}
\end{align}
where $n$ measures the staggered distance from the background  charge, with $n \in \{-L+1,-L+2,\ldots,L-1\}$, and $d \in \{1,2,\ldots,N-n-1\}$.
This pool satisfies the desired symmetry constraints, 
as $e^{i \theta \hat{O}_{mh}}$ is real and conserves charge.
The two terms in the RHS of the second line of Eq.~\eqref{eq:statPool} commute, and their exponentials can be converted to circuits without Trotter errors.

The convergence of the SC-ADAPT-VQE prepared ground state  $\vert \psi_{\text{ansatz}} \rangle $ 
to the true ground state  can be quantified with the deviation in the energy of the 
ansatz state $E_{\text{ans}}$ compared to the true  ground state  energy $E_{\text{gs}}$,
\begin{align}
\delta E = \frac{E_{\text{gs}} - E_{\text{ans}}}{E_{\text{gs}}} \ ,
\label{eq:EnergyDeviation}
\end{align}
as well as the infidelity density of the ansatz wavefunction with respect to the exact 
ground state,\footnote{
The average infidelity density is not an optimal metric to use as it asymptotes to 
the infidelity of the vacuum prepared with $\hat{U}^{\text{aVQE}}$ for large system sizes,
i.e., the deviation from the vacuum infidelity density will scale as $1/L$.
A better measure of infidelity would be the overlap of partially-reduced density matrices over a region of the lattice localized about the background charge.
Even with this, requiring a precision exceeding that of the vacuum state is not helpful.
}
\begin{align}
{\cal I}_L =\frac{1}{L}\left (1 - \vert \langle \psi_{\text{ansatz}} \vert  \psi_{\text{vac}} \rangle_{Q^+_{L-1}} \vert^2  \right )\ .
\label{eq:Infidelity}
\end{align}
The deviation in the energy and infidelity obtained from performing SC-ADAPT-VQE for $m=0.1$, $g=0.8$, and $L=\{8,10,12\}$ are given in Table~\ref{tab:AdaptEF}. 
The number of steps used to prepare the ground state  (and its convergence) can be found in App.~\ref{app:SCADAPTvac}.
The sequence of operators and the corresponding variational parameters are given in Table~\ref{tab:AdaptAng}.
It is surprising the first set of operators that are chosen have $d=4$, indicating that correlations of separation 4 are more important than separation 2, which come later in the ansatz.
One explanation is that the initial state has already included some of the short-range correlations (the vacuum prepared with $\hat{U}^{\text{avQE}}$  has $d=1$ and $d=3$ correlations). 
The initial state, $\vert \psi_{\text{init}} \rangle$ in Eq.~\eqref{eq:StatInit} is labelled as step 0 in Table~\ref{tab:AdaptEF}, and already has pretty good overlap with the desired state. 
After 4 steps, a deviation of the energy density of $\delta E \approx 0.012$ is reached which is sufficiently converged for our purposes.
The operators that are chosen, e.g., $\hat{O}_{mh}(3,4)$ and $\hat{O}_{mh}(1,4)$ in steps 1 and 2, and $\hat{O}_{mh}(3,2)$ and $\hat{O}_{mh}(-1,2)$ in steps 3 and 4, are related by a reflection about the position of the background  charge.
The optimal variational parameters are equal with
opposite signs, and the wavefunction that is being 
established has a version of the CP symmetry, but which is broken by boundary effects beyond 4 steps of SC-ADAPT-VQE.
The operator sequence is stable with increasing $L$, and the variational parameters are converging rapidly. 
This indicates that the extrapolation and scaling of these state preparation circuits should be robust.

Due to CP symmetry, the circuits for establishing the vacuum with a negative  background charge at site $L$ are identical to those for a positive charge, but with the variational parameters negated.
This technique can be generalized to prepare the ground state  in the presence of multiple background  charges located within the simulation volume.
As long as background  charges are well separated from each other and the boundaries, then the circuits (presented in the following section) simply need to be repeated around the location of each additional background  charge.
\begin{table}[!t]
\renewcommand{\arraystretch}{1.4}
\begin{tabularx}{\textwidth}{|c || Y | Y | Y | Y | Y || Y |  Y| Y | Y | Y |}
 \hline
  & \multicolumn{5}{c||}{$\delta E $ } & \multicolumn{5}{c|}{${\cal I}_L $} \\
  \hline
 \diagbox[height=23pt]{$L$}{\text{step}} & 0 & 1 & 2 & 3 & 4 & 0 & 1 & 2 & 3 & 4 \\
 \hline\hline
 8 & 0.0412 & 0.0356 & 0.0300 & 0.0242 & 0.0184 & 0.0164 & 0.0136 & 0.0109 & 0.0085 & 0.0062 \\
 \hline
 10 & 0.0317 & 0.0275 & 0.0232 & 0.0188 & 0.0145 & 0.0131 & 0.0109 & 0.0087 & 0.0069 & 0.0050 \\
 \hline
 12 & 0.0259 & 0.0226 & 0.0192 & 0.0157 & 0.0122& 0.0110 & 
0.0092 & 0.0073 & 0.0058 & 0.0043 \\
 \hline
\end{tabularx}
\caption{
The deviation in the energy $\delta E$, and infidelity density ${\cal I}_L$, of the SC-ADAPT-VQE prepared ground state  in the presence of a background  charge $\hat{Q}_{L-1} = +1$.
Results are shown through 4 steps of SC-ADAPT-VQE for $L=8,10,12$ and $m=0.1,g=0.8$. 
Step 0 corresponds to the initial state $\vert \psi_{\text{init}}\rangle$, defined in Eq.~\eqref{eq:StatInit}.}
 \label{tab:AdaptEF}
\end{table}
\begin{table}[!t]
\renewcommand{\arraystretch}{1.4}
\begin{tabularx}{0.5\textwidth}{|c || Y | Y | Y | Y |}
  \hline
 \diagbox[height=23pt]{$L$}{$\theta_i$} & $\hat{O}_{mh}(3,4)$ & $\hat{O}_{mh}(1,4)$ &  $\hat{O}_{mh}(3,2)$ &  $\hat{O}_{mh}(-1,2)$  \\
 \hline\hline
 8 & 0.1375 & -0.1375 & -0.1409 &  0.1409 \\
 \hline
 10 &  0.1367 & -0.1367  & -0.1404 & 0.1404 \\
 \hline
 12 & 0.1363  &  -0.1363  &  -0.1400 & 0.1400 \\
 \hline
\end{tabularx}
\caption{
The order of the operators $\hat{O}$ and variational parameters $\theta_i$ used to prepare the SC-ADAPT-VQE ground state  in the presence of a background  charge $\hat{Q}_{L-1} = +1$.
Results are shown through 4 steps of SC-ADAPT-VQE for $L=8,10,12$ and $m=0.1,g=0.8$. }
 \label{tab:AdaptAng}
\end{table}
%

\subsection{Quantum Circuits and Resource requirements}
\label{sec:Qcirqs}
\begin{figure}[!t]
    \centering
    \includegraphics[width=0.7\textwidth]{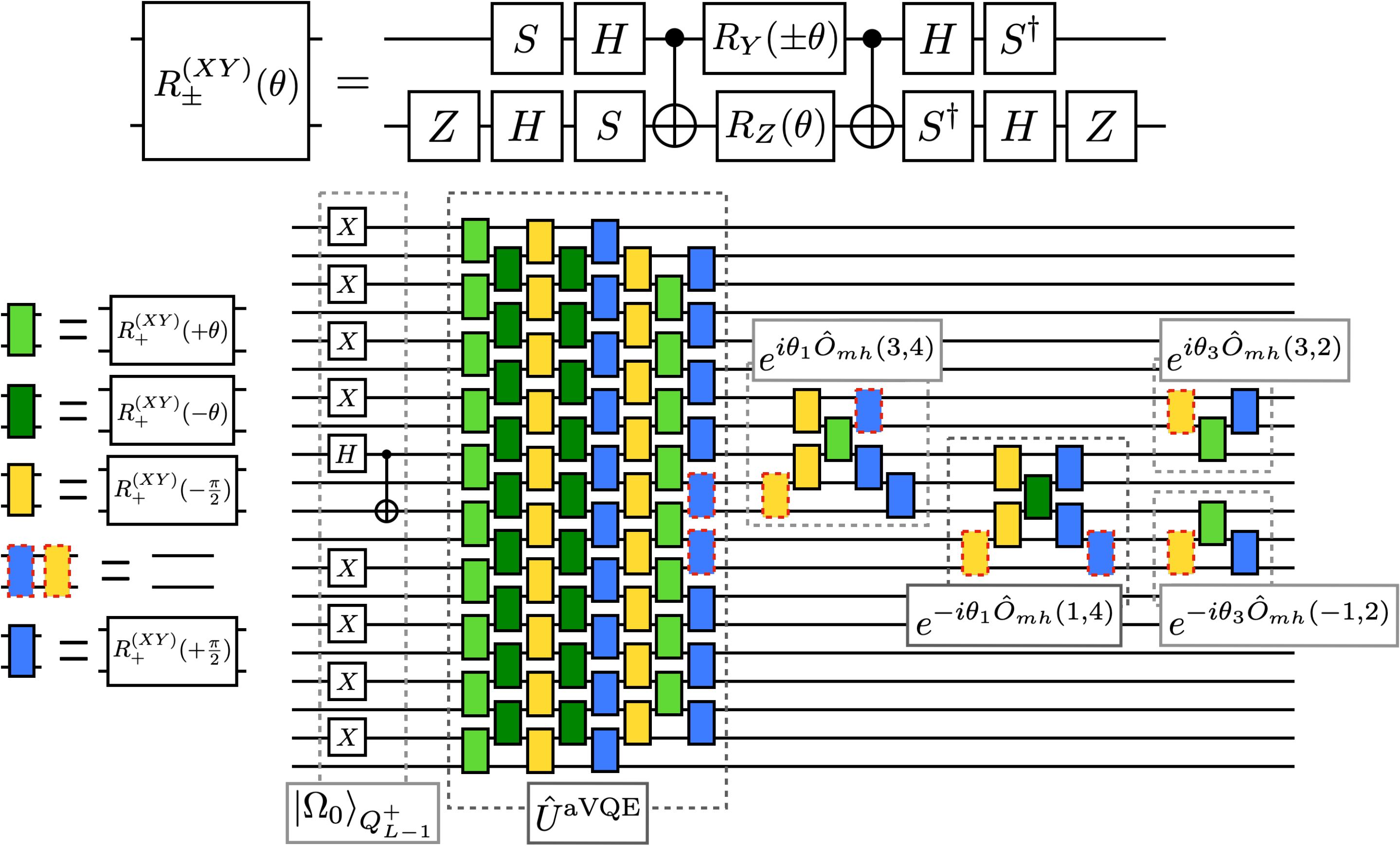}
    \caption{
    Circuits that prepare $\vert \psi_{\text{vac}}\rangle_{Q^+_{L-1}}$ with $L=10$ and a 
    background  charge at $Q_{L-1}=+1$. Initially, the strong coupling vacuum in the presence of the background charge $\vert \Omega_0 \rangle_{Q^+_{L-1}}$ is prepared. Next, the circuits that prepare the two-step SC-ADAPT-VQE vacuum without background  charges $\hat{U}^{\text{aVQE}}$, is applied. 
    Lastly, the four-step SC-ADAPT-VQE circuits for modifying the wavefunction around the position of the background  charge are applied.
    The parameters $\theta_i$ for $\hat{U}^{\text{aVQE}}$ are 
    given
    in App.~\ref{app:SCADAPTvac} and the angles
    for the second application of SC-ADAPT-VQE are given in Table~\ref{tab:AdaptAng}.
    The colored rectangles defining the circuits are defined in the left and top of the figure, with $R_{\pm}^{(XY)}(\theta) \equiv e^{-i \frac{\theta}{2}\left (\hat{Y}\hat{X} \pm \hat{X} \hat{Y} \right )}$. 
    }
    \label{fig:VacPrepCirc}
\end{figure}

\noindent
In the previous section, a sequence of 
unitary operations
that prepares the ground state  in the presence of a single background  charge was presented.
To perform simulations on a quantum computer, these 
unitary operations  
must be converted to a sequence of gates. 
Our circuit design is tailored toward  devices with linear nearest-neighbor connectivity, such as is native on IBM's quantum computers~\cite{ibmquantum},
and aims to minimize the circuit depth and two-qubit gate count.
The 
unitary operators forming
the operator pool in Eq.~\eqref{eq:statPool} are of the form $e^{i \theta\left ( \hat{Y}\hat{Z}^{d-1} \hat{X} - \hat{X}\hat{Z}^{d-1} \hat{Y}\right )}$,\footnote{The convention is that the operator on the far left acts on the lower numbered qubit, e.g. $\hat{Y}\hat{X} = \hat{Y}_n \hat{X}_{n+1}$.} 
and we will use the circuit  design introduced in our recent work~\cite{Farrell:2023fgd,Farrell:2024fit} that extends the techniques in Ref.~\cite{Algaba:2023enr}.
These circuits have an ``X'' shape, and are arranged in such a way to cancel the maximum number of CNOT gates.
An example of the circuit that prepares the ground state  in the presence of a background
charge $Q_{L-1}=+1$ for $L=10$ is shown in Fig.~\ref{fig:VacPrepCirc}.
This circuit has been decomposed into three parts. 
First, the strong coupling vacuum in the presence of the heavy charge $\vert \Omega_0\rangle_{Q^+_{L-1}}$ is prepared.
Next, the circuits that prepare the two step SC-ADAPT-VQE vacuum without a background charge are applied. 
These circuits are collectively denoted as $\hat{U}^{\text{aVQE}}$, and were treated in detail in Ref.~\cite{Farrell:2023fgd}.
Lastly, the circuits that implement the four step SC-ADAPT-VQE unitaries $e^{i \theta_i \hat{O}_i}$ in Sec.~\ref{sec:StatVac} are applied.
These circuits are localized and only modify the wavefunction around the position of the 
background charge.

As discussed in the previous section, preparing the ground state  in the presence of multiple 
background  charges is a straightforward extension of these circuits, provided the charges are well separated from each other and the boundaries. 
The first modification is to the preparation of $\vert \Omega_0\rangle_Q$: there is a hadamard-CNOT sequence centered around each heavy charge.
The second change is that the $e^{i \theta \hat{O}}$ are repeated around the center of each additional background  charge.
In total, the resources required for this state preparation are
\begin{align}
\text{\# of CNOTs = } 16L-12+25N_Q \ , \quad \text{CNOT depth = }35 \ ,
\end{align}
where $N_Q$ is the number of background  charges.
This circuit depth is well within the capabilities of current devices.
Note that the number of SC-ADAPT-VQE steps to maintain a constant quality of the prepared state will scale linearly with the confinement length $\xi$.
The circuit depth for each step of SC-ADAPT-VQE also scales with $\xi$: as ${\cal O}(\xi^2)$ for devices with nearest-neighbor connectivity and ${\cal O}(\xi)$ for devices with all-to-all connectivity.
As a result, for devices with nearest-neighbor connectivity, the circuit depth is expected to scale as ${\cal O}(\xi^3)$ for state preparation.

Once the initial state is prepared, time evolution can be implemented with the time-dependent Hamiltonian defined by the classical trajectory of the heavy charge, $Q(t)$ in Eq.~\eqref{eq:xva_softer}.
As shown in our previous work, time evolution in systems without background  charges can be reproduced up to exponentially small errors using a truncated electric interaction~\cite{Farrell:2024fit}. 
In Sec.~\ref{sec:SM}, it was argued that a similar procedure should also be possible for systems with heavy charges, provided the charge operators are suitably averaged over the extent of the heavy hadrons.
To get an estimate of the scaling, we assume a truncation of interactions between spatial charges separated by more than $\overline{\lambda} \approx \xi/2$  spatial sites. 
The resources required for one second-order Trotter step of time evolution can be estimated using the circuits in Ref.~\cite{Farrell:2024fit}, with a CNOT gate count of, 
\begin{equation}
\text{\# of CNOTs = }4(2L-1)+(2L-4\overline{\lambda})(\overline{\lambda}+1)(2\overline{\lambda}+1) -(L-2\overline{\lambda}+2) \ .
\end{equation}
Taking $\overline{\lambda} \sim \xi$, this gives a scaling of ${\mathcal O}(L \, \xi^2)$ for the number of two-qubit gates and a corresponding circuit depth of $\mathcal{O}(\xi^2)$.
A proper determination of the minimum $\overline{\lambda}$ required to reach a predetermined error threshold is left for future work.

To approach the continuum limit, $\xi$ is held fixed in physical units, while the lattice spacing is decreased, i.e., $\xi \sim a^{-1}$.
The number of Trotter steps must also grow as $\xi \sim a^{-1}$ 
as can be seen with the following argument. 
Trotterization of the kinetic term with a brickwork ordering only allows for correlations to spread two staggered sites per Trotter step.
Keeping the lattice volume traversed by the moving charge fixed in physical units implies that the number of staggered sites traversed scales as ${\mathcal O}(a^{-1})$.\footnote{It is possible that there is a Trotter ordering, different from brickwork, 
that improves this scaling.}
Therefore, the number of Trotter steps 
also scales as ${\mathcal O}(a^{-1})$ and time evolution is estimated to have a circuit depth that scales as ${\mathcal O}(a^{-3})$.
This is the same scaling as for the initial state preparation, giving a total circuit depth for simulating dynamics in dense matter in the Schwinger model to be ${\cal O}(a^{-3})$.
This depth would improve to ${\cal O}(a^{-2})$ on devices with all-to-all connectivity.
Of course, actual simulations that approach the continuum will need to be performed to validate these scaling arguments.

\section{Summary and Outlook}
\label{sec:Summary}
\noindent
The mechanisms responsible for energy-loss and transport in dense matter are key to understanding the evolution of
matter under extreme conditions: from high-energy collisions of large nuclei, to high-energy cosmic-ray penetrating ordinary matter, to the dynamics of core-collapse supernova.
There has been a long history of successfully using
classical techniques, such as Monte-Carlo simulation, to determine the 
electromagnetic responses when charged particles move through matter.
In contrast, 
the dynamics of high-energy quarks and gluons in dense matter is less understood, in part due to the non-perturbative phenomena of confinement and hadronization.
The ongoing heavy-ion experimental program, and the planned EIC at Brookhaven National Laboratory, 
will provide further experimental guidance in understanding energy-loss in dense matter, beyond the present capabilities of theoretical frameworks.
With an eye toward eventually performing quantum simulations of such processes in QCD,
we have performed real-time simulations of energy-loss and hadronization in 
the Schwinger model, with a simple model of a dense medium.
In particular, we have performed classical simulations of 
heavy-hadrons moving through regions of dense matter characterized by static heavy-hadrons.
These simulations have provided insight into internal excitations of hadrons, and the crucial role of quantum coherence between the particles that make up the dense medium.
The effects of quantum coherence between the constituents of matter are visible 
in the energy-loss as a function of incident velocity
in the highest density systems we have prepared.  
By subtracting the individual contributions, the remaining energy loss is attributed 
to quantum correlations  in the matter wavefunction with increasing density.
Further, we have provided scalable quantum circuits for preparing ground states with a finite density of heavy hadrons.
In combination with the time evolution circuits presented in our previous work~\cite{Farrell:2024fit}, we estimated the circuit depths required for large-scale quantum simulations of energy loss in the 
Schwinger model.
The outlook looks promising, and simulations of the dynamics of dense matter in the Schwinger model will be possible in the near-term. While the simulations we have performed in this work are limited to 1+1D QED, they are extendable to non-Abelian lattice gauge theories, such as QCD, and to higher numbers of spatial dimensions.  
They also constitute the leading order terms in systematic expansions, such as HQET and non-relativistic QCD (NRQCD). However, as we
have used classical sources to define the structure of the light degrees of freedom, further source-structure development
for nuclei and dense matter is required.

While it is no surprise, 
present-day simulations are significantly effected by relatively large lattice spacings,
restricted by the number of qubits (or qudits) that can be assembled into a quantum register 
to form a spatial lattice volume
that is large enough to contain more than a few confinement length scales.
The hadronic wavefunctions
have support only over a few lattice sites, rendering an obvious discretization of their wavefunctions, with large lattice spacing artifacts.  
One consequence is that when hadrons pass each other, there are relatively large differences between the incident and outgoing fields
at the lattice spacing scale.
In addition, the dispersion relation is such that the velocity of momentum modes has a maximum that is less than the speed of light. 
This occurs around the scale of the inverse lattice spacing 
(depending on mass and electric charge), and causes high-velocity hadrons 
to partially disintegrate as they move, even in the vacuum,
leaving behind a wake of low-energy hadronic excitations. 
This corresponds to fragmentation at fixed velocity, and is entirely a lattice spacing artifact.
These effects are mitigated by forming differences between propagation in matter and vacuum, but nonetheless present an unwelcome background from which to extract the physical fragmentation and hadronization.
These differences have a well-defined continuum limit, reflecting the target physics observables, 
and quantum simulations using multiple lattice spacings, tuned to known physics observables, 
are required
to make robust predictions with a complete quantification of uncertainties. 
An important result to highlight is that while lattice 
discretization effects are seen (and understood) in the energy loss of a single heavy-hadron moving through the vacuum, new effects are seen in the modification of the entanglement structure. This is pointing in the direction that quantum correlations are more sensitive to lattice artifacts than classical correlations.

On top of the discussion in the previous paragraph, the impact of lattice-spacing artifacts in high-energy processes cannot be under-estimated.  
One concern is that when colliding high-energy wavepackets together, the non-zero lattice spacing will induce scattering and fragmentation through the modified dispersion relation and beyond.
Care must be taken in such quantum simulations to ensure that the observed inelasticities are coming from physics, and not from the underlying lattice upon which the simulation is being performed.
Alternative formulations to Kogut-Susskind  where discretization errors are suppressed, 
such as improved-KS~\cite{Carlsson:2001wp,Carena:2022kpg,Dempsey:2022nys,Gustafson:2023aai} or improved-Wilson~\cite{Zache:2018jbt,Mathis:2020fuo,Mazzola:2021hma,Hayata:2023skf} 
Hamiltonians, are starting to be pursued. More development is required, leveraging knowledge from classical Euclidean lattice QCD
calculations.

A limitation of working in one spatial dimension is that peripheral collisions are not possible, and all collisions between the constituent electrons and positrons (partons)
that make up the hadrons are ``head-on''.  
In addition, there are no soft momentum-transfer processes, like bremsstrahlung radiation, due to the absence of dynamical gauge fields and the finite spatial extent of the lattice.
More realistic simulations of QCD will require advancing from a $U(1)$ to a $SU(3)$ lattice gauge theory, first in 1+1D and then in higher dimensions.
Development of  these more realistic simulations are underway, and will enable a study of the explicit role of 
non-Abelian color charges in the dynamics of dense matter.

\begin{acknowledgements}
\noindent
Roland would like to thank ECT* for support at the Workshop ``The physics of strongly interacting matter: neutron stars, cold atomic gases and related systems'' during which this work was completed.
We would also like to thank Mari Carmen Ba\~{n}uls for comments regarding tensor network simulations of the Schwinger model.
This work was supported, in part, by the U.S. Department of Energy grant DE-FG02-97ER-41014 (Roland), by U.S. Department of Energy, Office of Science, Office of Nuclear Physics, InQubator for Quantum Simulation (IQuS)\footnote{\url{https://iqus.uw.edu/}} under Award Number DOE (NP) Award DE-SC0020970 via the program on Quantum Horizons: QIS Research and Innovation for Nuclear Science\footnote{\url{https://science.osti.gov/np/Research/Quantum-Information-Science}} 
(Roland, Martin), 
the Quantum Science Center (QSC)\footnote{\url{https://qscience.org}} which is a National Quantum Information Science Research Center of the U.S.\ Department of Energy (DOE) (Marc).
This work is also supported, in part, through the Department of Physics\footnote{\url{https://phys.washington.edu}} and the College of Arts and Sciences\footnote{\url{https://www.artsci.washington.edu}} at the University of Washington.
This work was enabled, in part, by the use of advanced computational, storage and networking infrastructure provided by the Hyak supercomputer system at the University of Washington.\footnote{\url{https://itconnect.uw.edu/research/hpc}}
This research was done using services provided by the OSG Consortium~\cite{osg07,osg09,osgweb,osgweb2}, which is supported by the National Science Foundation awards \#2030508 and \#1836650.
This research used resources of the National Energy Research Scientific Computing Center (NERSC), a Department of Energy Office of Science User Facility using NERSC award NP-ERCAP0027114.
We have made extensive use of Wolfram {\tt Mathematica}~\cite{Mathematica}, {\tt python}~\cite{python3,Hunter:2007}, {\tt julia}~\cite{Julia-2017}, and {\tt jupyter} notebooks~\cite{PER-GRA:2007} in the {\tt Conda} environment~\cite{anaconda}, together with the {\tt CUDA}~\cite{besard2018juliagpu,besard2019prototyping}, {\tt ExponentialUtilities}~\cite{ExponentialUtilities}, and {\tt Expokit}~\cite{Expokit} {\tt julia} packages.
The numerical results that we have obtained
that support the findings of this study are available upon reasonable request.
\end{acknowledgements}

\clearpage
\appendix

\section{Spin Hamiltonian with External Charges}
\label{app:hamZ}
\noindent
The electric part of the Hamiltonian in Eq.~\eqref{eq:Hgf} can be expanded as,
\begin{equation}
    \frac{2}{g^2}\hat{H}_{el} = \sum_{j=0}^{2L-2}\bigg (\sum_{k\leq j} \hat q_k +Q_k\bigg )^2 = \sum_{j=0}^{2L-2}\bigg (\sum_{k\leq j} \hat q_k \bigg )^2 + 2 \sum_{j=0}^{2L-2} \bigg( \sum_{k\leq j}  \hat q_k   \bigg) \bigg( \sum_{l\leq j}  Q_l   \bigg) + \sum_{j=0}^{2L-2}\ \bigg( \sum_{k\leq j}  Q_k   \bigg)^2 \ ,
    \label{eq:backQ}
\end{equation}
The first term is unaffected by the presence of external charges and is given by,
\begin{equation}
\sum_{j=0}^{2L-2}\bigg (\sum_{k\leq j} \hat q_k \bigg )^2 = \frac{L^2}{2}+\frac{1}{4}\sum_{j=0}^{2L-2}\left(2L-j-\frac{1}{2}[1+(-1)^{j+1}]\right)\hat{Z}_j+\sum_{j=0}^{2L-3}\sum_{k=j+1}^{2L-2}\frac{2L-1-k}{2}\hat{Z}_j\hat{Z}_k \ .
\end{equation}
The terms that couple to the external charge are 
\begin{align}
2 \sum_{j=0}^{2L-2} \bigg( \sum_{k\leq j}  \hat q_k   \bigg) 
&\bigg( \sum_{l\leq j}  Q_l   \bigg)
\ +\ 
\sum_{j=0}^{2L-2}\ \bigg( \sum_{k\leq j}  Q_k   \bigg)^2 \nonumber \\
& = \sum_{j=0}^{2L-2}\left[ \bigg( \sum_{k\leq j}  Q_k   \bigg)^2 - \bigg( \sum_{l\leq j}  (-1)^l   \bigg)\bigg( \sum_{k\leq j}  Q_k   \bigg)\right] - \sum_{j=0}^{2L-2}\bigg(\sum_{l=j}^{2L-2}\sum_{m\leq l}Q_m\bigg) \hat{Z}_j \ ,
\end{align}
and contains terms proportional to the identity
and single $\hat{Z}_j$.

\FloatBarrier
\section{SC-ADAPT-VQE for Preparation of the Vacuum without Static Charges}
\label{app:SCADAPTvac}
\noindent
This appendix provides an overview of
the use of SC-ADAPT-VQE to prepare the vacuum without background charges.
The operator pool is constrained by the symmetries and conserved charges of the Schwinger model vacuum $\vert \psi_{\text{vac}}\rangle$: total charge, CP, and time reversal.
There is also an approximate translational symmetry in the volume 
for $L \gg \xi$, that is broken by the boundaries. 
This motivates organizing the pool into volume operators, $\hat{O}^V$, that are translationally invariant, and surface operators, $\hat{O}^S$, whose support is restricted to the boundary.
For the range of $m$ and $g$ explored in Ref.~\cite{Farrell:2023fgd} (including $m=0.1,g=0.8$), an effective pool of operators was found to be,
\begin{align}
\label{eq:poolComm}
\{ \hat{O} \}_{\text{vac}} &= \{ \hat O_{mh}^{V}(d) \ , \ \hat O_{mh}^{S}(0,d) \ , \ \hat O_{mh}^{S}(1,d)
 \}
\ ,\\
 \hat O_{mh}^{V}(d) & \equiv i\left [\hat{\Theta}_m^V, \hat{\Theta}_{h}^V(d)\right ]
 = 
\frac{1}{2}\sum_{n=0}^{2L-1-d}(-1)^n\left (
\hat X_n\hat Z^{d-1}\hat Y_{n+d} - 
\hat Y_n\hat Z^{d-1}\hat X_{n+d} 
\right )
\ ,\nonumber\\
\hat O_{mh}^{S}(0,d) & \equiv i\left [\hat{\Theta}_{m}^S(0), \hat{\Theta}_{h}^V(d) \right ]
 = 
\frac{1}{4}\left (\hat X_0\hat Z^{d-1}\hat Y_{d} - \hat Y_0\hat Z^{d-1}\hat X_{d} 
- \hat Y_{2L-1-d}\hat Z^{d-1}\hat X_{2L-1} + \hat X_{2L-1-d}\hat Z^{d-1}\hat Y_{2L-1}\right ) 
\ ,\nonumber \\
 \hat O_{mh}^{S}(1,d) & \equiv i\left [\hat{\Theta}_{m}^S(1), \hat{\Theta}_{h}^S(d) \right ]
 = 
\frac{1}{4}\left (\hat Y_1\hat Z^{d-1}\hat X_{d+1} - \hat X_1\hat Z^{d-1}\hat Y_{d+1} 
+ \hat Y_{2L-2-d}\hat Z^{d-1}\hat X_{2L-2} - \hat X_{2L-2-d}\hat Z^{d-1}\hat Y_{2L-2} \right )
\ . \nonumber
\end{align}
Time reversal invariance implies that the wavefunction is real, and constrains the pool operators to be  imaginary and anti-symmetric, e.g., $i$ times the commutators of orthogonal  operators $\hat{\Theta}$.
Here, $\hat{\Theta}_m^V$ ($\hat{\Theta}_m^S$) 
is a  volume (surface) 
mass term 
and $\hat{\Theta}_{h}^V(d)$ ($\hat{\Theta}_{h}^S(d)$) is a generalized volume (surface) hopping term that spans an odd number of fermion sites, $d$.
Only $d$ odd is kept as $d$ even breaks CP.\footnote{For $\hat{O}_{mh}^V(d)$ and $\hat{O}_{mh}^S(0,d)$, the range of $d$ is $d\in \{1,3,\ldots 2L-3\}$, and for $\hat{O}_{mh}^S(1,d)$ it is $d\in \{1,3,\ldots 2L-5\}$.}
The individual terms in each operator do not all commute, and they are converted to gates through a first-order Trotterization, ${\exp}(i \theta_i \hat{O}_i) \rightarrow \prod_j \hat{U}_j^{(i)}$, introducing (higher-order) systematic deviations from the target unitary operator.
The initial state for this application of SC-ADAPT-VQE is the strong-coupling vacuum without background  charges $|\Omega_0\rangle$.

The convergence of this algorithm and workflow was studied in detail in Ref.~\cite{Farrell:2023fgd} as a function of the number of SC-ADAPT-VQE steps. 
For $m=0.1,g=0.8$, it was found that  two steps of SC-ADAPT-VQE, was sufficient to achieve an infidelity density of ${\cal I}_L \approx 0.004$ and a deviation in the energy density of $\delta E \approx 0.006$.
These quantities are defined in Eq.~\eqref{eq:Infidelity} and Eq.~\eqref{eq:EnergyDeviation}, respectively.
The two-step vacuum preparation is used as a proof of principle for this work, and defines the vacuum state preparation unitary $\hat{U}^{\text{aVQE}}$ used in Sec.~\ref{sec:StatVac} to prepare the vacuum with background charges.
The operator sequencing and corresponding variational parameters for $m=0.1,g=0.8$, and several different $L$, are given in Table~\ref{tab:AdaptAngNoStat}.

\begin{table}[!t]
\renewcommand{\arraystretch}{1.4}
\begin{tabularx}{0.5\textwidth}{|c || Y | Y |}
  \hline
 \diagbox[height=23pt]{$L$}{$\theta_i$} & $\hat{O}_{mh}^V(1)$ & $\hat{O}_{mh}^V(3)$ \\
 \hline\hline
 10 & 0.3902 &  -0.0676  \\
 \hline
 11 & 0.3896 &  -0.0671  \\
 \hline
 12 & 0.3892 &  -0.0667  \\
 \hline
 13 & 0.3888 &  -0.0664  \\
 \hline
 14 & 0.3884 &  -0.0662  \\
 \hline
 \hline
 $\infty$ & 0.387 & -0.065 \\
 \hline
\end{tabularx}
\caption{
The order of the operators $\hat{O}$ and variational parameters $\theta_i$ used to prepare the two-step SC-ADAPT-VQE vacuum without background  charges.
Shown are results for $L=10-14$ and $m=0.1,g=0.8$. 
The extrapolation to $L=\infty$ is 
performed using the results from 
$L=11-14$, assuming an exponential dependence on $L$; 
see Appendix E of Ref.~\cite{Farrell:2023fgd} for details.}
 \label{tab:AdaptAngNoStat}
\end{table}
%

\bibliography{bib}

\end{document}